\documentclass[useAMS,usenatbib,usegraphicx]{mn2e}

\title[Pulsar Population]{Populating the Galaxy with pulsars I: stellar \& binary evolution}
\author[P.D. Kiel, J.R. Hurley, M. Bailes and J.R. Murray]{Paul D. Kiel$^{1}$\thanks{E-mail:
pkiel@astro.swin.edu.au (PDK)}, Jarrod R. Hurley$^{1}$, Matthew Bailes$^{1}$ and James R. Murray$^{1}$\\
$^{1}$Centre for Astrophysics and Supercomputing, Swinburne University of Technology, Hawthorn, Victoria, 3122, Australia}

\begin{document}

\date{Accepted xxx. Received xxx; in original form xxx}

\pagerange{\pageref{firstpage}--\pageref{lastpage}} \pubyear{2006}

\maketitle

\label{firstpage}

\begin{abstract}
The computation of theoretical pulsar populations has been a 
major component of pulsar studies since the 1970s.
However, the majority of pulsar population synthesis has only 
regarded isolated pulsar evolution.
Those that have examined pulsar evolution within binary systems 
tend to either treat binary evolution poorly or evolve the pulsar 
population in an ad-hoc manner.
Thus no complete and direct comparison with observations of the 
pulsar population within the Galactic disk has been possible to date.
Described here is the first component of what will be a complete 
synthetic pulsar population survey code.
This component is used to evolve both isolated and binary pulsars.
Synthetic observational surveys can then be performed on this 
population for a variety of radio telescopes.
The final tool used for completing this work will be a code comprised of 
three components: stellar/binary evolution, Galactic kinematics 
and survey selection effects.
Results provided here support the need for further (apparent) pulsar 
magnetic field decay during accretion, while they conversely suggest the 
need for a re-evaluation of the assumed \textit{typical} MSP formation 
process.
Results also focus on reproducing the observed $P\dot{P}$ diagram for 
Galactic pulsars and how this precludes short timescales for standard 
pulsar exponential magnetic field decay.
Finally, comparisons of bulk pulsar population characteristics are made 
to observations displaying the predictive power of this code, while
we also show that under standard binary evolutionary assumption 
binary pulsars may accrete much mass.
\end{abstract}

\begin{keywords}
binaries: close -- stars: evolution -- stars: pulsar -- stars: neutron -- 
Galaxy: stellar content
\end{keywords}

\section{Introduction}
\label{s:intro}

Since the tentative suggestion that neutron stars (NSs) form from 
violent supernova (SN) events (Baade \& Zwicky 1934a, 1934b) and the 
discovery of pulsars by Hewish et al. (1968) the number of observed pulsars 
has risen dramatically.
Surveys for radio pulsars have discovered over 1500 objects including
a rich harvest of binary and millisecond pulsars 
(Manchester et al.  2005; Burgay et al. 2006).
Precision timing of pulsars in binary systems has not only allowed precise 
tests of theories of relativistic gravity (e.g. van Straten et al. 2001), 
but also given insights into their masses and the nature of the binary 
systems they inhabit (for example Verbiest et al. 2007; 
Bell, Bailes \& Bessell, 1993). 
Of the first 300 or so pulsars to be discovered, only 3 were members
of binary systems, despite the progenitor population having an
extremely high binary fraction ($>$ 50\%; Duquennoy \& Mayor, 1991). 
This paucity of binary pulsars was an important clue about the origin and 
evolution of pulsars. 
Clearly something about pulsar generation was contributing to the disruption
of binary systems. 
We now believe that the supernovae in which pulsars are produced impart 
significant kicks to the pulsars, that make their survival prospects 
within binary systems bleak. 

Today, there are well over 100 binary pulsars known, and their spin periods
and inferred magnetic field strengths offer the opportunity to attempt
models of binary evolution and pulsar spin-up that explain their 
distribution in the pulsar magnetic field-spin period 
($B_{\rm s}$-$P$) diagram. 
To do this properly, one should take models of an initial population of 
zero-age main-sequence (ZAMS) single stars and binaries, trace the binary 
and stellar evolution, including neutron star spin-up effects, calculate 
their Galactic trajectories and initial distribution in the Galaxy, and 
then perform synthetic surveys assuming a pulsar luminosity and beaming 
function.
This is what we wish to achieve.
The large number of assumptions that we require to complete this 
effort caution against the absolute predictive power of such a model. 
For instance, it is easy to demonstrate that trying to use such a model 
to predict something like the merger rate of NS-black hole (BH), 
or double pulsars/NSs based solely on a consideration of the total 
number and mass distributions of un-evolved ZAMS binaries would be folly. 
However, the relative numbers of two or more populations can often 
only depend upon relatively few model assumptions 
(as shown in HTP02; Belczynski, Kalogera \& Bulik 2002; 
O'Shaughnessy et al. 2008).
With enough observables in time we might hope to build up a self-consistent 
theory of binary and pulsar evolution. 
This paper, the first in a series, aims to address the above suggested 
binary and stellar pulsar evolution population synthesis component.
Later work will combine this product with the kinematic and selection effect
components, facilitating direct comparison of theory with observations.
Therefore this paper is only a first step towards such a complete 
description but one that, as we will show, can already constrain models 
of neutron star magnetic field evolution and spin-up.

\begin{figure}
  \includegraphics[width=84mm]{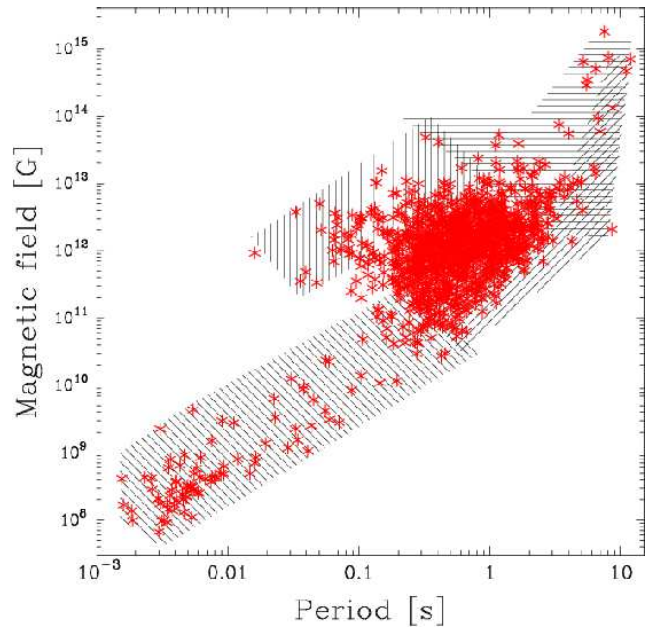}
  \caption{
  Magnetic field vs spin period of observed pulsars 
  within the Galaxy (stars, including some within globular clusters).
  The observations are overlaid with a cartoon depiction 
  of the reason for the particulars of the pulsar parameter 
  space distribution. 
  The area covered by vertical bars primarily 
  arises from the canonical pulsar birth properties 
  and spin-down due to magneto-rotational energy 
  losses.
  The horizontal barred regions are formed for 
  the most part by deviation of pulsar birth 
  properties from the average values from whence 
  pulsar spin-down evolution commences.
  Forward leaning horizontal bars depict the 
  region in which a luminosity law or loss of obliquity 
  of beam direction (or a combination of these) induces 
  a decrease of numbers in the observed population.
  The final observed pulsar region arises from 
  binary evolution, although a number of these 
  systems are isolated it is possible they were 
  formed in binaries which disrupted due to the 
  explosion or ablation of the secondary star.
  \label{f:BvsP}}
\end{figure}

As alluded to above, observations of pulsars take the form of spin 
measurements -- both the spin period $P$ and spin period derivative 
$\dot{P}$ -- in which the surface magnetic field $B_{\rm s}$ 
and characteristic age of the pulsars are inferred 
(details are discussed when we introduce our model of 
pulsar evolution).
It is instructive to plot $\dot{P}$ vs $P$ (hereafter $P\dot{P}$) and 
$B_{\rm s}$ vs $P$ diagrams and any theoretical model 
must be able to reproduce these if it is to be successful.
In Figure~\ref{f:BvsP} we show the $B_{\rm s}$ vs $P$ diagram
of $\sim1400$ pulsars taken from the ATNF Pulsar 
Catalogue\footnote{http://www.atnf.csiro.au/research/pulsar/psrcat/} 
(Manchester et al. 2005).
We see three distinct regions of the parameter space being populated.
A large island of relatively slow spinning pulsars with high surface 
magnetic fields (and thus high $\dot{P}$s) is joined via a thin bridge of 
pulsars to another, smaller, island of relatively rapid rotators with 
low magnetic fields.

The present theoretical explanation for the distinct groups 
of observed pulsars is rotating magnetic neutron stars that are either 
isolated or reside in a binary system (see Wheeler 1966; Pacini 1967; 
Gold, 1968; Ostriker \& Gunn 1969; Gunn \& Ostriker 1970; 
Goldreich \& Julian 1969, van den Heuvel 1984,
Colpi et al. 2001; Bhattacharya 2002; 
Harding \& Lai 2006).
It is those NSs that evolve within a binary system that may attain the 
low surface magnetic fields and low rotational periods, 
i.e. rapid rotators, as mentioned above.
The double pulsar binary J0737-3039A \& B (Burgay et al. 2003; 
Lorimer 2004; Lyne et al. 2004; Dewi \& van den Heuvel 2004) shows the 
distinction between rapid rotators and slow rotators clearly.
Here we have a binary system consisting of a millisecond pulsar (MSP: 
$P = 22.7$ ms and $B_{\rm s} = 7\times10^{9}$ G) and its `standard'
pulsar companion ($P = 2.77$ s and $B_{\rm s} = 6\times10^{12}$ G).
Although it seems likely that it is the process of accretion onto the 
NS that induces NS magnetic field decay (or apparent magnetic field decay)
this evolutionary phase is still highly contentious and theories abound on 
how the decay may occur. 
One such theoretical argument is of accretion-induced field decay 
via ohmic dissipation of the accreting NSs crustal currents.
This is due to the heating of the crust which in turn increases 
the resistance in the crust 
(Konar \& Bhattacharya 1997, 1999a, 1999b; Geppert \& Urpin 1994; 
Urpin \& Geppert 1995; Romani 1990; Urpin, Geppert \& Konenkov 1997).
An alternative explanation of accretion-induced field decay consists of 
screening or burying the magnetic field with the accreted material 
(Lovelace, Romanova \& Bisnovatyi-Kogan 2005; Konar \& Choudhuri 2004; 
Choudhuri \& Konar 2002; Cumming, Zweibel \& Bildsten 2001; 
Melatos \& Phinney 2001; Bisnovatyi-Kogan \& Komberg 1974; 
Taam \& van den Heuvel 1986).
While yet another argument considers vortex-fluxoid (neutron-proton) 
interactions.
Here the neutron vortices latch on and then drag
the proton vortices (which bear the magnetic field) radially, 
the radial direction is induced by either spin-up (inwards) or spin-down 
(outwards) of the NS (Jahan-Miri 2000; Muslimov \& Tsygan 1985;
Ruderman 1991a, b \& c).

Below is a general overview of pulsar evolutionary paths.
A greater level of detail  
is given within the very informative reviews presented by
Colpi et al. (2001),  Bhattacharya (2002), 
Choudhuri \& Konar (2004), Payne \& Melatos (2004), Harding \& Lai (2006) 
and references therein.
There are a number of possible binary and stellar evolutionary 
paths one can conceive that allow the formation of a pulsar.
If the end product is an isolated `standard' pulsar,
the star may have always been isolated and evolved from
a massive enough progenitor star (mass greater than $\sim 10~M_\odot$) 
with no evolutionary perturbations from outside influences.
However, it may have been that the progenitor pulsar was bound
in an orbit and depending on the orbital parameters when the NS formed, 
mass loss and the velocity kick during the asymmetric SN event 
(Shklovskii 1970) could contrive to disrupt the binary, allowing us to 
observe a solitary pulsar.
Alternatively, if the secondary star was also massive
(but originally the less massive of the two stars) disruption may occur
in a second SN event.
The creation of a millisecond (re-cycled; Alpar et al. 1982) 
pulsar requires the accretion of material onto the surface 
of the NS at some point during its lifetime.
For this to occur the primary star is first required to form a NS and the 
binary must survive the associated SN event.
Then, some time later, the secondary star evolves to overflow its Roche-lobe
and initiate a steady mass transfer phase.
Any accretion of the donated material onto the NS will increase 
its spin angular momentum, thus spinning up the pulsar, potentially
to a millisecond spin period.
The companion of the resultant MSP will typically be a low-mass main sequence
star, a white dwarf (WD: a MSP-WD binary) or another pulsar.
Formation of a double pulsar binary is problematic because if
more than half of the systems mass is lost in the second SN event the
binary will be disrupted.
This is unless a kick is directed toward the vicinity of the companion 
MSP with just the right magnitude to overcome the energy change owing
to mass loss but not too strong as to disrupt the binary.
If instead the binary is disrupted by the SN then there would be both a 
solitary `standard' pulsar and a single MSP.
One other suggested method for producing a single millisecond pulsar is
for the donor star to be evaporated or ablated by the extremely active
pulsar radiation (which may be modelled in the form of a wind).
The pulsar is then said to be a black widow pulsar (van Paradijs et al. 1988).

As mentioned above, the theories of pulsar evolution can be tested in a
statistical manner by comparing observations to population
synthesis results.
This theoretical approach has been adopted previously for pulsars and 
other stellar systems (some examples are: Dewey \& Cordes 1987;
Bailes 1989; Rathnasree 1993;
Possenti et al. 1998; Portegies Zwart \& Yungelson 1998;
Possenti et al. 1999; Willems \& Kolb 2002; O'Shaughnessy et al. 2005; 
Kiel \& Hurley 2006; Dai, Liu \& Li 2006;
Story, Gonthier \& Harding 2007 and numerous works based on 
\textsc{StarTrack}, presented in Belczynski et al. 2008).
The level of detail in the population synthesis calculations 
varies, along with the methodologies the authors implemented.
For example, Bailes (1989) considered pulsar selection effects in 
some detail, however, only roughly considered binary evolutionary 
phases and the method in which this affects pulsar evolution.
Willems \& Kolb (2002) considered NS populations and how
these affect the resultant NS populations, however, they were not
able to directly compare with observations as no selection effects
were modelled.
In a slightly different approach (empirically based)  
Kim, Kalogera \& Lorimer (2003) estimated
the merger rate (via gravitational waves) of double NS (DNS) binaries  
within the Galaxy by selecting the physical \textit{observable} 
DNS pulsar properties from appropriate distribution 
functions for many pulsars and weighting these against the observed 
Galactic disk DNS pulsars PSR B1913+16 and PSR B1534+12.
Taking into account selection effects of the pulsar population
for large scale surveys and producing many pulsar models
they were able to give confidence levels for their merger rate 
estimates.
In comparison, we will select the \textit{initial evolutionary}
parameters from distribution functions and evolve all stars
forward in time from stellar birth on the ZAMS
through until the current time.
This gives us the ability to constrain many stellar and binary
evolutionary features while lending us the flexibility to, for example,
compare different populations of stars with each other and observations.
In this way we can aim to further constrain uncertain parameter values. 
However, even at this early stage we must place a strong 
word of caution regarding the issue of parameter variation
-- strong degeneracies can apply amongst parameters 
(e.g. O'Shaughnessy et al. 2005), where a succession of 
parameter changes can mask the effect of another, so extensive 
modelling is required before definite conclusions can be made.

\begin{figure}
  \includegraphics[width=84mm]{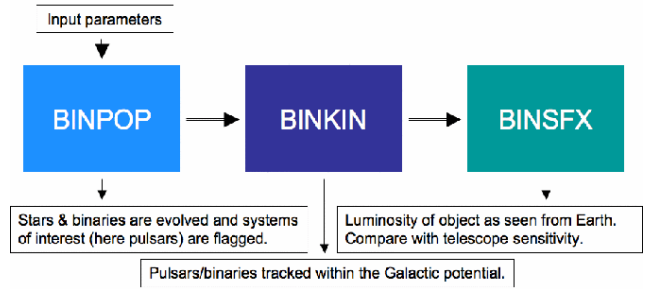}
  \caption{Flow chart of our synthetic pulsar population 
    survey process.
    The work presented here focuses on the first module,
    \textsc{binpop} (see text for details).
  \label{f:fchart}}
\end{figure}

Our goal is to create a generic code for producing 
synthetic Galactic populations.
This will comprise three modules: \textsc{binpop}, 
\textsc{binkin} and \textsc{binsfx} (see the flow 
chart in Figure~\ref{f:fchart} for a representation 
of how these fit together).
The first module, \textsc{binpop}, covers the stellar 
and binary evolution aspects and as such is a 
traditional population synthesis code in its own right.
The second module, \textsc{binkin}, follows the 
positions of both binary systems and single stars 
within the Galactic gravitational potential.
The third module imposes selection effects on the 
simulations, thus giving simulated data that can 
compare directly to observations.
This last consideration is in some regards the most 
important as without detailed modelling of selection 
effects any comparison of population synthesis 
simulations to observations is crude (Kalogera \& Webbink 1998).
This paper focuses on a description of the 
\textsc{binpop} module and, in particular,
the pulsar population that it produces.
We consider in depth modelling of pulsar evolution 
in terms of the spin period and the magnetic field 
coupled with mass accretion.
Simulating these processes will help in constraining 
the stellar birth properties of NSs and also lead 
to a greater understanding of aspects of NS evolution 
such as the formation event itself -- the supernova.
It will also allow predictions of the composition of the 
Galactic pulsar population.
Follow-up work will discus \textsc{binkin} with a 
focus on the kinematic evolution of the pulsar 
distribution within the Galaxy, and SN veolicty 
kicks -- and \textsc{binsfx}.

This paper is organised as follows.
Section 2 gives an overview of the rapid binary evolution
population synthesis code used in this research. 
This is followed in Section 3 by a detailed description 
of the pulsar modelling techniques that have been added 
to this code to create \textsc{binpop}.
Section 4 gives examples of the pulsar evolutionary
pathways that can be followed with \textsc{binpop} and how 
these can be affected by choices in the algorithm.
Population synthesis results in the form of $P\dot{P}$ 
comparisons are given in Section 5 followed by a 
discussion in Section 6.
In particular we wish to draw the readers attention to 
the results shown in Section~\ref{s:further} which extend 
beyond the basic $P\dot{P}$ description.

\section{Rapid binary evolution \& Population Synthesis}
\label{s:bmodel}

The research presented here makes use of the first 
module of our synthetic Galactic population code.
This is called \textsc{binpop} and  wraps the 
Hurley, Tout \& Pols (2002: HTP02) 
\textsc{binary stellar evolution} (\textsc{bse})
code\footnote{See also http://astronomy.swin.edu.au/$\sim$jhurley/
and relevant links.} 
(with updates as described in Section 3) within a 
population synthesis package.

The \textsc{bse} algorithm is described in detail 
by HTP02 and an overview is given in Kiel \& Hurley (2006).
The aim of \textsc{bse} is to allow rapid and robust, 
yet relatively detailed, binary evolution based on the 
most up-to-date  prescriptions/theories for the various 
physical processes and scenarios that are involved.
In its most basic form the \textsc{bse} algorithm can be 
thought of as evolving two stars forward in time
-- according to the \textsc{single star evolution (sse)} 
prescription described in Hurley, Pols \& Tout (2000)
-- while updating the orbital parameters.
After each time-step the algorithm checks whether either 
star has over-flowed its Roche-lobe and depending on the 
result the system is evolved accordingly, i.e. as a detached, 
semi-detached, or contact binary.
During these phases the total angular momentum of the 
system is conserved while orbital and spin changes owing 
to tides, mass/radius variations, magnetic braking and 
gravitational radiation are modelled.

Within \textsc{bse} an effort is made to model all 
relevant stellar and binary evolutionary processes, 
such as mass transfer and common-envelope (CE) evolution.
Invariably this involves making assumptions about how 
best to deal with elements of the evolution that are uncertain.
An example that is relevant to pulsar evolution is the choices 
made for the nature of the star produced as a result of the 
coalescence of two stars, where at least one NS is involved.
If the merger involves a NS and a non-degenerate star then a 
Thorne-\.{Z}ytkow object (T\.{Z}O: Thorne \& \.{Z}ytkow 1977)
is created.
Detailed modelling of these objects must deal with neutrino physics
and processes such as hypercritical accretion and currently 
the final outcome is uncertain 
(Fryer, Benz \& Herant 1996; Podsiadlowski 1996).
Possibilities include rapid ejection of the envelope
(the non-degenerate star) 
to leave a single NS that has not accreted any mass or collapse
of the merged object to a BH.
In \textsc{bse} the former is currently invoked -- an unstable T\.{Z}O.
On the other hand, the coalescence of a NS with a degenerate 
companion is assumed to produce a NS with the combined mass of the 
two stars, unless the companion is
a BH in which case the NS is absorbed into the BH.
For steady transfer of material onto a NS 
-- in a wind or via Roche-lobe overflow (RLOF) --
it is assumed in \textsc{bse} that the NS can 
accrete the material up to the Eddington limit (Cameron \& Mock 1967).
However, there is some uncertainty as to what extent this limit applies
as there may be cases where energy generated in excess of the limit
can be removed from the system (e.g. Beloborodov 1998).
For this reason the Eddington limit is included as an 
option in \textsc{bse} (HTP02).
Another area of uncertainty is what happens to a massive oxygen-neon-magnesium
WD when it accretes enough material to reach the Chandrasekhar limit
-- does it explode as a mass-less supernova or form a NS?
Currently \textsc{bse} follows the models of Nomoto \& Kondo (1991) 
which suggest that electron capture on ${\rm Mg}^{24}$ nuclei leads 
to an accretion-induced collapse (AIC) and a NS remnant (Michel 1987).
In the formation of a NS via the AIC method \textsc{bse} assumes 
that no velocity kick is imparted onto the NS while some 
other population synthesis works have assumed a small but non-zero
velocity kick for these NSs (Ivanova et al. 2007; 
Ferrario \& Wickramasinghe 2007).
These examples illustrate some of the decision making involved with creating
a prescription-based evolution algorithm and the interested reader is
directed to HTP02 for a full description as well as a list of options included
within the \textsc{bse} code.

Subsequent to HTP02 the following changes have been made to 
the \textsc{bse} algorithm:
\begin{itemize}
\item an option to calculate supernova remnant masses using the 
prescription given in Belczynski, Kalogera \& Bulik (2002) which 
accounts for the possibility of the fall-back of material during 
core-collapse supernovae (this is now the preferred 
option in \textsc{bse});
\item an algorithm to compute the stellar envelope structure 
parameter, $\lambda$, (required in CE calculations) from the results 
of detailed models (Pols, in preparation) -- previously this was set 
to a constant whereas models show that it varies with mass and evolution 
age (Dewi \& Tauris 2001; Podsiadlowski et al. 2003);
\item the addition of equations to detect, and account for, 
the existence of an accretion disk during mass transfer on to a 
compact object (following Ulrich \& Burger 1976); and,
\item an option to reduce the strength of winds from helium stars.
\end{itemize}
These changes are documented in greater detail by 
Kiel \& Hurley (2006).
Further additions to the \textsc{bse} algorithm relating 
to pulsar evolution will be described in the next section.

There are also a number of areas where the \textsc{bse} 
algorithm could be improved in the future.
The integration scheme used for the differential 
equations in \textsc{bse} (and for the equations introduced 
here) employs a simple Euler method (Press et al. 1992) 
with the time-step depending primarily on the nuclear evolution of the stars.
It has been suggested that this will lead to inaccurate results for the orbital
evolution, in particular when integrating the tidal equations 
(Belczynski et al. 2008), and an implementation of a higher-order 
scheme will be a priority for the next revision of the
\textsc{bse} code.
It will also be desirable to include a RLOF treatment for 
{\it eccentric} binaries and the
addition of an option for tidally-enhanced mass-loss from 
giant stars close to RLOF
(along the lines of Bona\v{c}i\'{c} Marinovi\'{c}, Glebbeek \& Pols 2008).
Another area of uncertainty is the strength of stellar winds 
from massive stars (Belczynski et al. 2008).
This is a feature which can be varied within \textsc{bse} 
but we do not utilise it or examine its
effect on compact object formation in this work.

Reflecting the variety of processes involved in binary 
evolution, and the uncertain nature of many of these, 
there are inevitably a substantial number of free 
parameters associated with a binary evolution code. 
Unless otherwise specified we assume the standard choices
for the \textsc{bse} free parameters as listed in Table 3 of HTP02.
This includes setting $\alpha_{\rm CE} = 3$ for the CE efficiency parameter.
In addition we use the variable $\lambda$ for CE evolution
and the remnant mass relation of Belczynski et al. (2002),
as described above.
We also use $\mu_{\rm nHe} = 0.5$ in the expression for
the helium star wind strength (see Kiel \& Hurley 2006 for details)
and we impose the Eddington limit for accretion onto remnants.
Where applicable (Sections 4 and 5) we will take care 
to note our choices for any \textsc{bse} parameters that we vary.

Our aim is to produce a Galactic population of pulsars.
We do this using the statistical approach of population synthesis.
The first step is to produce realistic initial populations of single
and binary stars.
For each initial binary the primary mass, secondary mass and orbital 
separation are drawn at random from distributions based on observations
(see Section~\ref{s:method} for details).
For single stars only the stellar mass is required for each star.
We then set a random birth age for each binary (or star) and follow the 
evolution to a set observational time.
By selecting those systems that evolve to become pulsars we
can produce a $P$ vs $\dot{P}$ diagram for the Galactic pulsar population
-- assuming (at least initially) that all generated pulsars are observable. 
This is repeated for a variety of models and comparisons made to 
observations.
This is the method we use here.
However, it is possible to determine the initial parameters 
from a set grid and then convolve the results with the above 
mentioned initial distribution functions to determine if the 
binary contributes to the Galactic population.
For more information on the grid method see Kiel \& Hurley (2006).

\section{Modelling a Pulsar}
\label{s:Modelling}

We now describe our methods for evolving pulsars --
both isolated and within a binary system.
This requires a number of additions to the \textsc{bse} algorithm.
We point out here that although most important phases of stellar
and binary evolution are modelled we still draw the initial pulsar 
evolutionary parameters -- spin period and magnetic field -- 
from distributions. 
This is the method employed in all previous pulsar population 
synthesis simulations.
The need for this arises because there is no complete
theoretical method for modelling the SN event (Janka et al. 2006), 
and thus no strong link between the pre-SN star and post-SN star.
Aspects such as the initial NS mass, however, are better determined.
Also, kick velocities imparted to NSs at birth are generally 
selected from distributions that match observations of NS space 
velocities.
We investigate the initial NS spin period and magnetic field 
distributions and extend previous studies by linking these to 
this kick velocity (see Sections 3.4 and 5.5).
We also note that although some previous pulsar population synthesis
studies take into account selection effects to allow direct 
comparison with observations we do not do this here in any detail.
As a first step we do consider beaming effects but mainly we leave 
this for future work.
Our wish at this stage is to test that our evolution algorithm can 
populate the required regions of the observed parameter space.
Considering direct number 
and birth rate comparisons with observations will follow when the full 
three module code is complete.

\subsection{Previous efforts}
There have been many attempts at modelling NS evolution.
Some, as given in the introduction, try to generalise or 
parametrise the evolution to allow rapid computation for 
statistical considerations.
Another method, however, is to perform detailed modelling of a 
single NS. 
Debate on NS evolution has concentrated mainly on the 
structure of NSs and the evolution of the magnetic field.
Most statistical studies have considered only those NSs that are 
isolated pulsars and here concern is, generally, on the magnetic
field decay timescale or space velocity and magnetic field correlation.
The semi-analytical studies of Gunn \& Ostriker (1970),
Phinney \& Blandford (1981), Vivekanand \& Narayan (1981)
and Lyne, Manchester \& Taylor (1985) found a magnetic field 
decay timescale of order a few million years.
In comparison, studies by Taylor \& Manchester (1977),
van den Heuvel (1984) and Stollman (1987) found the decay timescale 
to be $> 100$ Myr.
More recently population studies of pulsars have tended towards
a greater level of detail in the treatment of pulsar evolution and
modelling of selection effects.
Again, differing studies find a variation of magnetic field decay
time-scales.
Those such as Bailes (1989), Rathnasree (1993); Bhattacharya et al. (1992), 
Hartman et al. (1997), Lorimer et al. (1993), 
Lorimer, Bailes \& Harrison (1997), Regimbau \& de Freibas Pacheco (2000), 
Regimbau \& de Freibas Pacheco (2001), Dewi, Podsiadlowski \& Pols (2005) 
and Faucher-Giguere \& Kaspi (2006) conclude that field decay occurs 
on timescales comparable to the age of old pulsars.
While Gonthier et al. (2002) and Gonthier, Van Guilder \& Harding (2004) 
suggest NSs evolve with shorter decay time-scales ($\sim$ few Myr).
However, Faucher-Giguere \& Kaspi (2006) suggest this later result is 
an artifact of the pulsar radio luminosity law assumed by
Gonthier et. al. (2002) and Gonthier, Van Guilder \& Harding (2004).
The field decay time-scale is a parameter within our models that is 
allowed to vary.
As we will here, Dewey \& Cordes (1987), Bailes (1989), Rathnasree (1993), 
Dewi, Podsiadlowski \& Pols (2005) and Dai, Liu \& Li (2006) take into 
consideration pulsars that may evolve within binary systems and follow 
the orbital properties along with the pulsar spin and magnetic field.
Previously either the treatment of binary evolution
or pulsar evolution has been limited, although the work of
Dai, Liu \& Li (2006) utilises the \textsc{bse} and a 
limited form of pulsar physics.
No published work to date has fully considered the effect of magnetic 
field decay within both single and binary pulsar evolution in such detail 
as presented here.

\subsection{Single or wide binary pulsar evolution}
We first consider the evolution of a NS that does not accrete any material,
either an isolated NS or one in a wide enough orbit to allow single 
star-like evolution.
Although, in the latter case we note that aspects of binary evolution such as 
tidal interaction are followed in step with the pulsar spin evolution.
We evolve a solitary pulsar by assuming a pulsar magnetic-braking model 
of the form,
\begin{equation}
\label{e:pdot}
\dot{P} = K \frac{B_{\rm s}^2}{P},
\end{equation}
(Ostriker \& Gunn 1969) with $\dot{P}~[s/s]$, $B_{\rm s}~[G]$, $P~[s]$ and $
K \sim 6.3\times10^{-54}~[R_\odot^3sM_\odot^{-1}]$.
Here it is assumed that the beam is $90^\circ$ to the pulsar equator.
This is of course a simplification because the beam must be off-axis 
to pulse (Spitkovsky 2006).
Other methods of pulsar spin evolution may be included in future 
iterations of the model.
Some more examples are magnetar spin-down evolution (Harding, 
Contopoulos \& Kazanaz 1999) and spin-down of oblique  
(Contopoulos \& Spitkovsky, 2006; Spitkovsky 2006)
 or perpendicular rotators (Goldreich \& Julian 1969).
The magnetic field is also assumed to evolve in time due 
to ohmic dipole decay which, in its simplest form, may be modelled as,
\begin{equation}
\label{e:bsol}
B_{\rm s} = B_{\rm s0}\exp\left(-\frac{T-t_{\rm acc}}{\tau_{\rm B}}\right),
\end{equation}
where $B_{\rm s0}$ is the initial NS surface magnetic field, T is the age
of the pulsar, $\tau_{\rm B}$ is the magnetic field 
decay time-scale and $t_{\rm acc}$ 
is the time the pulsar has spent accreting matter via RLOF.
Allowing the magnetic field to decay is inspired by observations, 
where older isolated pulsars are shown to have lower magnetic 
fields than younger pulsars and is now a standard evolutionary model
within the literature (Gunn \& Ostriker 1969; Stollmen 1987).
This apparent magnetic field decay is still observed even when 
detailed modelling of selection effects are taken into account
(Faucher-Giguere \& Kaspi 2006), though as mentioned above the
timescale varies due to assumptions used within the modelling.
To integrate these equations into the \textsc{bse} we must follow the
angular momentum of the NS using the magnetic-braking model 
of Equation~\ref{e:pdot}.
Differentiating the angular velocity with time and applying 
the chain rule we have the angular velocity time derivative,
\begin{equation}
\label{e:omegadot1}
\dot{\Omega} = -\frac{\Omega}{P}\dot{P}
\end{equation}
Substituting Equation~\ref{e:pdot} into Equation~\ref{e:omegadot1}
allows us to find the angular velocity rate of change (here it is 
a spin-down effect) in terms of the intrinsic features of the pulsar,
\begin{eqnarray}
\dot{\Omega} & = & -K\frac{\Omega}{P^2}B_{\rm s}^2\\
             & = & -\frac{K}{4\pi^2}\Omega^3 B_{\rm s}^2\\
\label{e:omegadot2}
             & = & -K_2\Omega^3 B_{\rm s}^2
\end{eqnarray}
where $K_2 \sim 2.5\times10^{-52}$.
It is Equation~\ref{e:omegadot2} which we use directly with the code,
evolving the NS magnetic-braking angular momentum, $J_{\rm mb}$ with,
\begin{equation}
\dot{J}_{\rm mb} = \frac{2}{5}MR^2\dot{\Omega},
\end{equation}
which is the angular momentum equation for a solid spherical rotator 
with radius $R$, mass $M$ and angular velocity $\Omega$.
This is then removed from the total NS angular momentum, the system is 
updated (this includes the NS spin-down being transfered to 
the orbital angular momentum via tides) and we step forward in time.

In the work presented here we are evolving the surface magnetic field,
$B_{\rm s}$ of the NS.
Strictly speaking if we were to compare our theoretically calculated
magnetic field to observations we should use the magnetic field strength
at the light cylinder radius -- that is the radius at which if the magnetic
field is considered a solid rotator it is rotating at the speed of light.
When observations of the spin period and spin period derivative are made it
is the magnetic field at this radius that is being sampled.
When we calculate the period derivative this assumed change of magnetic 
field is taken into account and as such we have two reasons for comparing 
our theoretical models to observations in the period-period derivative 
parameter space.
Firstly, it is $\dot{P}$ that is actually being measured observationally 
and secondly the $\dot{P}$ calculation naturally takes into account the 
light cylinder magnetic field (as it makes use of the pulsar magnetic moment).

\subsection{Pulsar evolution during accretion}
\label{s:Paccev}

Allowing the decay of a NSs magnetic field during an accretion 
event is believed to be the cause of the observed pulsar $P\dot{P}$ 
distribution (see Section~\ref{s:intro}).
Until now this has not been tested with any detailed 
modelling of binary, stellar and pulsar evolutionary physics. 
During any accretion phase angular momentum is transfered to 
the accreting object and it is spun-up.
To follow this evolution we use the standard equations 
describing the system angular momentum as given in HTP02.
While steady mass transfer occurs we do not allow any decrease in
the NS rotational velocity due to the magnetic-braking process 
described by Equation~\ref{e:omegadot2}.
Instead the NS angular momentum increases while the magnetic 
field decays exponentially with the amount of mass accreted,
\begin{equation}
\label{e:bacc}
B_{\rm s} = B_{\rm s0}\exp \left(-k\frac{\Delta M}{M_\odot}\right),
\end{equation}
where k is a scaling parameter that determines the rate of decay.
Although there is no physical basis for Equation~\ref{e:bacc} if
accretion onto a NS does cause -- even at the very least -- an apparent decay 
of the NS magnetic field then the process must be tied, in some 
manner, to the accreted mass (see Section~\ref{s:intro}).
A similar method of magnetic field evolution during accretion was 
first suggested by Shibazaki et al. (1989) and Romani (1990), 
with their equation of,
\begin{equation}
\label{e:bshi}
B_{\rm s} = \frac{B_{\rm s0}}{1+\frac{\Delta M}{M_{\star}}},
\end{equation}
which is comparable to our Equation~\ref{e:bacc}.
$M_{\star}$ is a parameter that Shibazaki et al. (1989) 
fit to observations and found to be of order 
$10^{-4}~M_\odot$.
Similar to Shibazaki et al. (1989) we do not assume any particular
model of field decay during the accretion phase 
(i.e. burial of field lines, see Section~\ref{s:intro}), only that field 
decay following Equation~\ref{e:bacc} does occur.

Typical accretion onto a NS has the in-falling matter interacting 
with the magnetosphere.
Once the magnetic pressure dominates the gas pressure the matter
is channelled along magnetic field lines to be accreted onto the 
magnetic poles (Pringle \& Rees 1972).
Not all mass transfer events within a binary system, however, 
result in accretion of material.
One such example with much relevance here is the `propeller' 
(Illarionov \& Sunyaev 1975) phase (Davidson \& Ostriker 1973; 
Illarionov \& Sunyaev 1975; Kundt 1976; Savonije \& van den Heuvel 1977).
Here we have a NS with a sufficiently large magnetic field and 
rotational velocity.
The mass being transfered from the companion falls towards the NS
but instead of following the magnetic field lines to the surface 
the material is ejected from the system by the rapidly rotating 
magnetic field lines.
The magnetic field acts as a propeller, changing rotational energy of the NS
into kinetic energy for the now outgoing matter.
This phase has a noticeable effect on the NS.
Due to the loss of rotational energy by the NS it spins down.
This is therefore an important evolutionary phase to consider 
during accretion and we have implemented a method for modelling 
it within \textsc{bse}.
We follow Jahan-Miri \& Bhattacharya (1994) who
allow the difference between the Keplerian angular velocity, 
$\Omega_{\rm K}$, at the pulsar magnetic radius $R_{\rm M}$ and 
the co-rotation 
angular velocity, $\Omega_*$, 
($V_{\rm diff} = \Omega_K \left(R_{\rm M} \right) - \Omega_*$) 
to decide whether the accretion phase spins the NS up or down.
The magnetic radius is assumed to be half the Alfven radius, $R_{\rm m}$,
the radius at which the magnetic pressure balances the ram-pressure, 
\begin{equation}
\frac{B_{\rm s}^2R^6}{8\pi R_{\rm m}^6} =  \frac{\left(2GM\right)\dot{M}}
        {4\pi R_{\rm m}^{5/2}}.
\end{equation}
This gives
\begin{equation}
R_{\rm m} = 3.4\times 10^{-4} R_\odot \left( 
      \frac{B^2_{\rm s} R^6}{M^{1/2}\dot{M}} \right)^{2/7}.
\end{equation}
The rate of change of the angular momentum for an accreting NS is 
then (cf. Equation 2 Jahan-Miri \& Bhattacharya 1994),
\begin{equation}
\dot{J}_{\rm acc} = \epsilon V_{\rm diff} R_{\rm M} \dot{M},
\end{equation}
where $\epsilon$ is a parameter that allows for any uncertainties
in the efficiency of coupling  the magnetic field and matter.
During wind accretion $\epsilon$ is considered unity, however,
when Roche-lobe mass transfer occurs we allow $\epsilon$ to 
vary slightly with the mass transfer rate, that is 
$\epsilon \sim 10^{-14} / \dot{M}$ (see Jahan-Miri \& Bhattacharya 1994).
Jahan-Miri \& Bhattacharya (1994) conclude that this method is 
relatively robust given much of the physics of accretion 
and capture flow are ignored.
A similar though slightly more detailed method is used 
within the pulsar population synthesis 
work of Possenti et al. (1998).
During the propeller phase we assume that the entire discarded matter 
leaves the binary system.

Something that has not received a large amount of attention in
binary NS population synthesis work is the assumed amount of angular 
momentum gained by a NS when accreting mass via a wind from its companion
(see however the informative works of Liu \& Li 2006; Dai, Liu \& Li 2006).
Because the spin histories of NSs are important for our comparisons to 
observations, angular momentum accretion is something that we feel should
be examined.
In particular this evolutionary phase may be important when modelling the
apparent bridge of observed pulsars that connect the standard pulsar
island to the recycled island.
To examine this effect we multiply the angular momentum of the accreted mass
by the efficiency parameter $\Xi_{\rm wind}$ giving,
\begin{equation}
\Delta J_{\rm wind} = \frac{2}{3} \Xi_{\rm wind} \Delta M R_2^2 \Omega_2,
\end{equation}
where $J_{\rm wind}$ is the angular momentum transferred in 
the wind, $\Delta M$ is the accreted mass, $R_2$ is the companion 
stellar radius and $\Omega_2$ is the companion star angular velocity.
Basing the variability of $\Xi_{\rm wind}$ on the assumption that the NS
magnetic field plays an integral part when NS angular momentum accretion
occurs, we allow $\Xi_{\rm wind}$ to vary as,
\begin{equation}
\Xi_{\rm wind} = MIN\left(1,0.01\left(2\times10^{11}/B_{\rm s}\right)+0.01\right).
\end{equation}
By assuming the above form of $\Xi_{\rm wind}$ we allow 
larger magnetic fields to dominate the flow of angular momentum, 
while the simplicity depicts our lack of knowledge of the physical 
processes in such an evolutionary phase.
An example of the uncertainty is shown in the work of Ruffert (1999) who
estimates an angular momentum accretion efficiency between $0.01$ and $0.7$
(hence our lower limit for $\Xi_{\rm wind}$).

\subsection{Initial pulsar parameter selection}
\label{s:InitPparams}
As mentioned earlier previous pulsar population synthesis 
work begin by selecting the initial period, magnetic field, 
mass, etc, of the pulsar from distributions, ignoring any 
earlier evolutionary stages the star may have passed through.
If the distributions for each parameter are realistic this method
is considered robust to first order.
This is the method followed here for the initial spin period of the pulsar,
$P_0$, and the initial surface magnetic field, $B_{\rm s0}$.
We select from flat distributions between 
$P_{\rm 0min}$ and $P_{\rm 0max}$ for the initial spin period 
and between $B_{\rm s0min}$ and $B_{\rm s0max}$ for the initial magnetic field.
Typical values for these parameters are $P_{\rm 0min} = 0.01\,$s, 
$P_{\rm 0max} = 0.1\,$s, $B_{\rm s0min}= 10^{12}\,$G and 
$B_{\rm s0max}= 3 \times 10^{13}\,$G, although
these can vary (see Table~\ref{t:table1}).

In actual fact the exact initial pulsar spin period and magnetic field
distributions are unknown.
However, Spruit \& Phinney (1998) have previously suggested that
there is a connection between the three pulsar birth characteristics:
imparted velocity, spin period and magnetic field strength.
A parameter linked to the SN event and one in which we claim to have
some form of observational contraint is $\textbf{V}_{\rm kick}$.
The velocity kick is randomly selected from a Maxwellian 
distribution with a dispersion of $V_{\sigma}$
(as in HTP02 and Kiel \& Hurley 2006).
With this in mind we trial a method in which the initial pulsar
birth parameters $P_0$ and $B_{\rm s0}$ are linked to 
$\textbf{V}_{\rm kick}$.
In this `SN-link' method we have,
\begin{equation}
\label{e:Pinit2}
P_0 = P_{\rm 0min} + P_{\rm 0av}\left( V_{\rm kick}/V_{\sigma} \right)^{\rm n_{\rm p}},
\end{equation}
and 
\begin{equation}
\label{e:Binit1}
B_{\rm s0} = B_{\rm s0min} + B_{\rm s0av}\left( V_{\rm kick}/V_{\sigma} \right)^{\rm n_{\rm b}}\quad,
\end{equation}
where $P_{\rm 0av}$ is the average initial period and $B_{\rm s0av}$ 
is the average initial magnetic field.
The exponents $n_{\rm p}$ and $n_{\rm b}$ allow us to parameterise 
the effect the SN event has on the initial pulsar period and magnetic 
field but we take both equal to one throughout this work.
An assumption here is that a more energetic explosion
would be able to impart a greater initial angular velocity onto 
the proto-NS than a lesser explosion, which may in turn introduce a more 
effective dynamo action increasing the initial magnetic field strength.

We note here that due to the lack of a complete understanding 
of SNe, we are at present not attempting to find the true pulsar 
initial period and magnetic field distributions {\it but} trying 
to depict how modifying the initial pulsar birth region in the 
$P\dot{P}$ distribution affects the final pulsar $P\dot{P}$ distribution.
Previous population synthesis models have differed on their preferred
initial parameter distributions (compare Faucher-Giguere \& Kaspi 2006
and Gonthier, Van Guilder \& Harding 2004; Story, Gonthier \& Harding 2007), 
however, any complete pulsar population synthesis must be able to 
reproduce the observed pulsars associated with SN remnants.

\subsection{Electron capture supernova}
\label{s:ecsn}

An important phase in NS evolution which we have touched 
on briefly already is the birth of NSs in SN events.
Previous numerical models of SNe suggested that convection may drive
the formation of assymetries in SNe (Herant, Benz \& Colgate 1992;
Herant et al. 1994).
More recently there have been exciting developments in this area
driven by the successful modelling of 
SN explosions in multi-dimensional (spatially) hydrodynamic 
simulations (Mezzacappa et al. 2007; 
Marek \& Janka 2007; Fryer \& Young 2007).
These models show high levels of convection due to the stationary 
accretion shock instability (found numerically by 
Blondin \& Mezzacappa 2003).
This instability leads to an asymmetry in the explosion event and 
provides a mechanism for delivering large SN velocity kicks.
Observationally, proper motion studies of pulsars have detected large 
isolated pulsar space velocities of order $1000$ km/s (e.g. 
Lyne \& Lorimer 1994).
Furthermore, mean pulsar space velocities appear to be in excess of the mean
for normal field stars (e.g. Hobbs et al. 2005).
These results show the necessity of imparting velocity kicks to NSs.
As such most previous population synthesis 
works have assumed a Maxwellian SN velocity kick distribution 
with $V_\sigma \sim 200 - 500$~km/s (see above in 
Section~\ref{s:InitPparams}).

Of late however, there has been a body of evidence that some 
NSs must have received small velocity kicks, of order $50$ km/s 
(Pfahl et al. 2002; Podsiadlowski et al. 2004).
Observationally Pfahl et al. (2002) found a new type of 
high-mass X-ray binary which exhibited low eccentricities and wide 
orbits, suggesting a low velocity kick imparted during the SN event.
This is backed up by evidence of the relatively large amount 
of NSs within globular clusters compared to the estimated number if 
one considers they receive an average velocity kick of
$V_\sigma> 200$ km/s (Pfahl, Rappaport \& Podsiadlowski 2002; 
and more recently in the theoretical studies of 
Kuranov \& Postnov 2006 and Ivanova et al. 2007).
The type of SN explosion that theoretically causes small 
velocity kicks, and is in vogue at present, is the electron capture 
(EC) method.
Basically this is the capture of electrons in the stellar 
core by the nuclei $^{24}Mg$, $^{24}Na$ and $^{20}Ne$ 
(Miyaji et al. 1980, see their Figure 8 and Table 1).
This results in a depletion 
of electron pressure in the core, facilitating the core collapse 
to nuclear densities if the final core mass is within the mass range of
$1.4$ to $2.5~M_\odot$ (Nomoto 1984; 1987).
The bounce of material due to the halt of the collapse by the 
strong force thus drives the traditional SN event.
Taking into account both single star evolution 
(Poelarends et al. 2007) and binary evolution 
(Podsiadlowski et al. 2004) suggests that the 
initial mass range of stars that may explode via the EC SN 
mechanism resides somewhere within $6 - 12~M_\odot$, with some 
dependence on metallicity and assumptions made in stellar modelling.
Early work on the EC SN explosions suggested that it is a 
prompt event, in which any asymmetries would not have time 
to occur.
Thus allowing the nascent NS a small SN kick velocity.
This is in contrast to SN explosions from more massive initial 
mass stars ($>12~M_\odot$) which produce
rather slow explosions (many hundreds of ms; Marek \& Janka 2007).
We note that the latest oxygen-neon-magnesium SN explosions 
(initial stellar mass range 
of $8 - 12~M_\odot$) by Kitaura, Janka \& Hillebrandt (2006) 
show that the EC SN is not so prompt as previously thought 
(few hundred ms) and also energy yields are a factor of 10 less.
However, these authors still find low recoil velocities for the nascent NS, 
again because hydrodynamic instabilities are unlikely to form.

As detailed in Ivanova et al. (2007) there are a number of 
stellar and binary evolutionary pathways which may lead to EC SN.
Although EC SN may occur from processes of single star 
evolution we concentrate here on a relatively newly 
recognised binary evolution example (Podsiadlowski et al. 2004).
A primary star with a zero-age-MS mass between 
$\sim 8 - 12~M_\odot$ that lives in the appropriate 
binary system may experience a mass transfer phase 
in which its convective envelope is stripped from it 
before the second dredge-up phase on the asymptotic giant branch.
This lack of second dredge-up leaves behind a more 
massive helium core than would otherwise have remained 
if the hydrogen-rich envelope had not been stripped 
off.
Yet it is a less massive core than is left behind 
for stars $> 12~M_\odot$ 
(see Podsiadlowski et al. 2004, Figure 1).
Depending upon the rate of mass loss via a wind the 
helium core may then evolve through the relevant phases
to explode as an EC SN (Podsiadlowski et al. 2004).
The ability for a binary system to produce a helium star
with a mass within the assumed range of $1.4$ to $2.5~M_\odot$ 
as given above is much greater than for a single star 
(Podsiadlowski et al. 2004).

The possibility of EC SNe is included as an option in \textsc{bse} 
for the following cases:
\begin{itemize}
\item a giant star with a degenerate oxygen-neon-magnesium 
core that reaches the Chandrasekhar mass -- for the stellar evolution
models used in \textsc{bse} this coresponds to an initial mass range of 
$\sim 6 - 8~M_\odot$ (for solar metalicity: HTP02);

\item a helium star with a mass between $1.4~M_\odot - 2.5~M_\odot$
(as set by Podsiadlowski et al. 2004 based on the work of Nomoto 1984; 1987).
\end{itemize}
When a NS forms in these cases we use an electron capture 
kick distribution with $V_\sigma = 20$ km/s (see NS kick distribution
of Kiel \& Hurley 2006).

\subsection{NS magnetic field lower limit}
\label{s:Bbot}
Most detailed magnetic field decay models show a `bottom' or lower limit 
value of the magnetic field in which no further decay is possible.
Here, this lower limit is a model parameter, $B_{\rm bot}$, and, as given 
in Zhang \& Kojima (2006), should be of order $10^7 - 10^8~\rm{G}$.
We use $5\times 10^7 \rm{G}$ throughout this work.

\subsection{Radio emission loss: The death line}
\label{s:death}

Generally, it is believed electron-positron pairs are the particles 
that, when accelerated within the magnetic field of a NS, produce the 
observed radio emission.
These pairs are assumed to form in the polar cap (PC) region
of the NS (PC model: Arons \& Scharlemann 1979; Harding \& Muslimov 1998) or 
in the outer NS magnetosphere beyond the polar caps
(as in the outer gap model: Chen \& Ruderman 1993).
Depending upon the chosen model it is possible to predict when or
under what circumstances a NS will turn off its beam mechanism and 
thus stop being observable as a pulsar.
These predictions form a region in the $P\dot{P}$ parameter space where
the NS is not able to sustain pair production.
The edge of this region is commonly referred to as the pulsar death line.
Until recently efforts within pulsar population synthesis 
calculations for simulating the death-line have only considered isolated 
pulsars.
Such models have been based on the work by Bhattacharya et al. (1992) 
who, following Ruderman \& Sutherland (1975), give the condition,
\begin{equation}
\label{e:death1}
\frac{B_{\rm s}}{P^2} > 0.17\times 10^{12} G/s^2.
\end{equation}
Zhang, Harding \& Muslimov (2000) extend the above equation to 
consider differing forms of particle scattering and acceleration methods 
but still only consider isolated pulsar evolution when constructing 
their pulsar death line models.
These conditions, however, do not consider the physics involved in rapidly 
rotating NSs, such as MSPs, and can not account for the observed pulsar 
cut off at lower periods.

Here we must turn to the work of Harding, Muslimov \& Zhang (2002, 
henceforth HMZ) who, when considering the PC model, make allowance for 
MSP spin periods and masses.
In the long line of work completed in this field (Harding \& Muslimov 
1998, 2001, 2002; HMZ) it is shown that there are a number of conditions 
which can be used to differentiate between the pair-production domain 
and the pair-production prohibited domain.
These conditions rely heavily on two factors, (1) the type of radiation 
that forms a cascade of the aforementioned pair-produced particles and
(2) the height at which this cascade develops -- the pair formation front.
For (1) HMZ calculate models based on three radiation dominant pair-production 
mechanisms.
They assume the pair-produced particles may be created by 
curvature radiation (CR), which itself is released from accelerated 
particles forced to move along curved magnetic field lines,
or via inverse Compton scattering (Hibschman \& Arons 2001; HMZ), 
that is, soft photons (i.e. relatively low energy photons) being 
scattered to high energies by accelerated high energy charged particles.
If a small number of these soft photons have energies near the 
cyclotron frequency they will scatter with a much greater cross section 
and thus provide a larger number of high energy photons 
(the inverse Compton photons).
Therefore, HMZ consider two inverse Compton models, resonant inverse Compton 
scattering and nonresonant inverse Compton scattering.
For (2), if the pair formation front occurs at an altitude less 
than the radius of the polar cap (defined by the strength of the 
magnetic field) the pulsar is said to be unsaturated.
If the pair formation front is at an altitude greater than the polar
cap radius the pulsar is in the saturated regime.
These regimes are important because they determine in what type of electric 
field the pair-produced particles are accerlate in.
Unsaturated pairs may accelerate in an increasing (in altitued) electric field 
(parallel to the magnetic field lines), while saturate pairs are 
accelerated in a constant electric field.

In this work, for simplicity, we limit ourselves to only consider 
the CR death line HMZ model.
Our slightly modified HMZ-CR death line equations are:

\begin{eqnarray}
P &<& P_*^{\rm CR}, \\
\log\left( \dot{P} \right) &=& \frac{21}{8}\log\left(P \right) - \frac{7}{4}\log\left(\bar{f}_{\rm prim}^{\rm min} \right)\\
& &-\frac{3}{2}\log\left( I \right) + 
 \frac{19}{2}\log\left( R \right) - 9.66
\end{eqnarray}

\begin{eqnarray}
P &>& P_* ^{\rm CR}\\
\log\left( \dot{p} \right) &=& \frac{5}{2}\log\left(P\right) - 2 \log\left( \bar{f}_{\rm prim}^{\rm min} \right)\\
&&- \frac{3}{2}\log\left( I \right) + 11\log\left( R \right) - 19.84,
\end{eqnarray}
where, $P < P_*^{\rm CR}$ is the unsaturate regime and 
$P_*^{\rm CR} = 4.64\times 10^{-7} B_{\rm s}^{4/9}$ is 
the CR saturated-unsaturated limit in seconds.
The modification simply allows the use of solar units for $M$ and $R$ -- $B$
is in Gauss, $P$ in seconds.
A strength of the HMZ death lines is the fact that stellar evolutionary 
features such as mass and radius play a role in how the radio emission is
formed.
In fact, HMZ show how important binary mass transfer is in modifying
the death line equations.
Thus one parameter included within these equations is the NS moment 
of inertia, $I$, which we wish to calculate.
To calculate $I$ we are required to assume an equation of state and
mass-radius relationship,
\begin{equation}
I = \frac{2}{7}\left(1-2.42\times10^{-6}\frac{M}{R}-2.9\times10^{-12}\frac{M^2}{R^2}\right)^{-1}MR^2
\end{equation}
and
\begin{equation}
R = 2.126\times10^{-5}M^{-1/3}.
\end{equation}
Here, I is taken from Tolman (1939) and as shown in Figure~6 of 
Lattimer \& Prakash (2001) this equation is a good fit to detailed models
of NSs for differing equation of states down to 
$\frac{M}{R} \sim 10^{-7}~M_\odot / R_\odot$.
HMZ show that the pulsar death line is particularly sensitive
to the assumed equation of state, any future work on this area
must take this into consideration.
After assuming an equation of state and mass-radius relation we may still
modify the effect of the HMZ death-line equations by varying
$\bar{f}^{\rm min}_{\rm prim}$, 
the minimum pulsar efficiency required for 
pair production to occur, a parameter used by HMZ to fit their 
parameterised equations to detailed model results.
Harding \& Muslimov (2002) suggest upper and lower bounds on what 
value $\bar{f}^{\rm min}_{\rm prim}$ should take for each model.
For CR these values are $0.1$ and $0.5$ for lower and higher respectively. 

\section{Some evolution examples}
\label{s:evex}
\begin{figure*}
  \includegraphics[width=168mm]{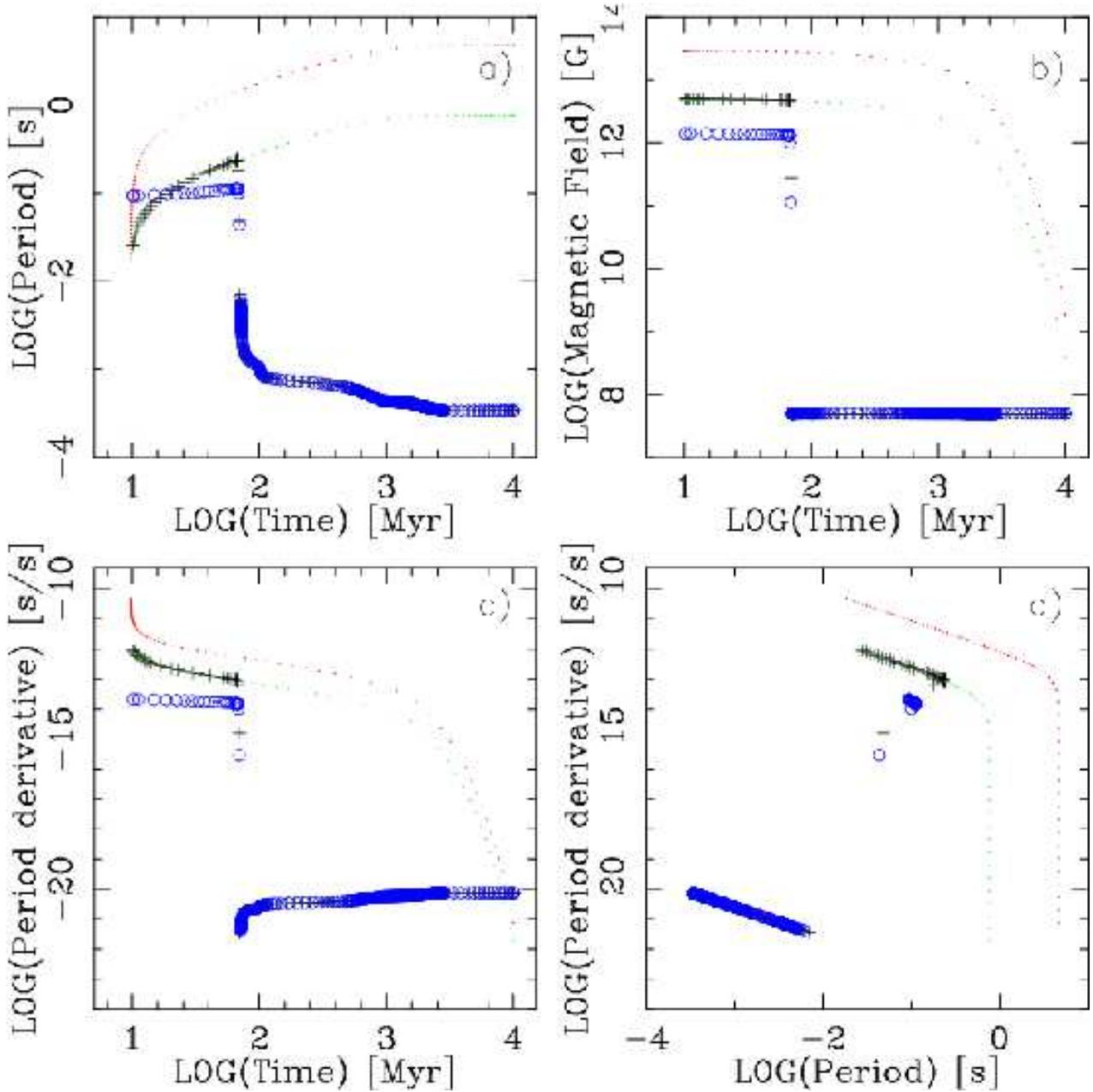}
  \caption{
  The evolution of pulsar particulars, $P$, $B_{\rm s}$ and $\dot{P}$
  for differing initial parameter $P$ and $B_{\rm s}$ values.
  There are two isolated pulsar evolutionary pathways depicted here,
  these are the dotted paths.
  Two binary pulsar evolutionary paths are also shown (pluses 
  and circles).
  See text for further details.
  \label{f:X1}}
\end{figure*}

To examine the newly implemented \textsc{bse} pulsar evolution
we now take a detailed look into the evolution of an isolated
pulsar and a binary system containing a pulsar.
This also allows us to investigate how sensitive pulsar evolution
is to uncertain parameters in the binary and pulsar model.

\subsection{An isolated pulsar}

We begin by considering an isolated $20~M_\odot$ star 
initially residing on the main-sequence (MS) with
solar metallicity, $Z = 0.02$.
This star very quickly becomes a $2.3~M_\odot$ NS;
spending only $8.8$ Myr as a MS star, 0.016 Myr on the 
Hertzsprung gap (HG), skipping the first giant branch to 
ignite core helium burning (which lasts for $1.0$ Myr) and 
then moving onto the first asymptotic giant branch, where 
after $0.02$ Myr it becomes a NS.
The NS, when born, is given a randomly selected spin period of 
$P_0 = 2\times10^{-2}~\rm{s}$ and a surface magnetic field of 
$B_{\rm s0} = 5.0\times10^{12}~\rm{G}$.
This means the NS in this example has a period derivative of 
$\dot{P}_0 = 8.5\times10^{-12}~\rm{s/s}$.
The subsequent evolution of the NS spin period, magnetic field and 
period derivative are shown in Figures~\ref{f:X1}a), b) and c)
respectively (the lower dotted curves in each plot).
The evolution in the $P\dot{P}$ diagram is shown in Figure~\ref{f:X1}d).

After $10~\rm{Gyr}$ the stellar spin period, magnetic field and 
period derivative are `observed' as, $0.76~\rm{s}$, 
$3.7\times10^8~\rm{G}$ and $2.04\times10^{-22}~\rm{s/s}$
respectively.
No pulsar has ever been observed with such parameters, 
so it is safe to assume that this NS, after $10~\rm{Gyr}$, 
would be beyond the pulsar death line.
In fact, if we assume a death line of the form given by 
Equation~\ref{e:death1} then the pulsar would die after a system time 
of $4555$ Myr.
This is the case if we assume $\tau_{\rm B} = 1000$ Myr.
However, if we assume instead that the magnetic 
field decays on a much faster time-scale, say, 
$\tau_{\rm B} = 5$ Myr, then the NS parameters evolve to a 
somewhat different configuration.
In this case the NS ends with $P = 6.9\times10^{-2}$ s, 
$B_{\rm s} = 5\times10^7$ G and $\dot{P} = 3.6\times10^{-23}$ s/s.
Here the NS magnetic field reaches the lower limit of 
$5\times10^7~\rm{G}$ in just over $600~\rm{Myr}$ which is about 
the same time that the pulsar would cross the death line.
If we assume the star has a metallicity of $0.001$ or $0.0001$ 
then a BH is born directly.
This is because the boundary between NS and BH formation 
(which is governed by the asymptotic giant branch core mass which collapses to 
a $3~M_\odot$ remnant) varies slightly with metallicity.
In terms of initial stellar mass this is about $21~M_\odot$ for solar 
metallicity whereas for a $Z = 0.0001$ population it is $19~M_\odot$.
\begin{figure*}
  \includegraphics[width=168mm]{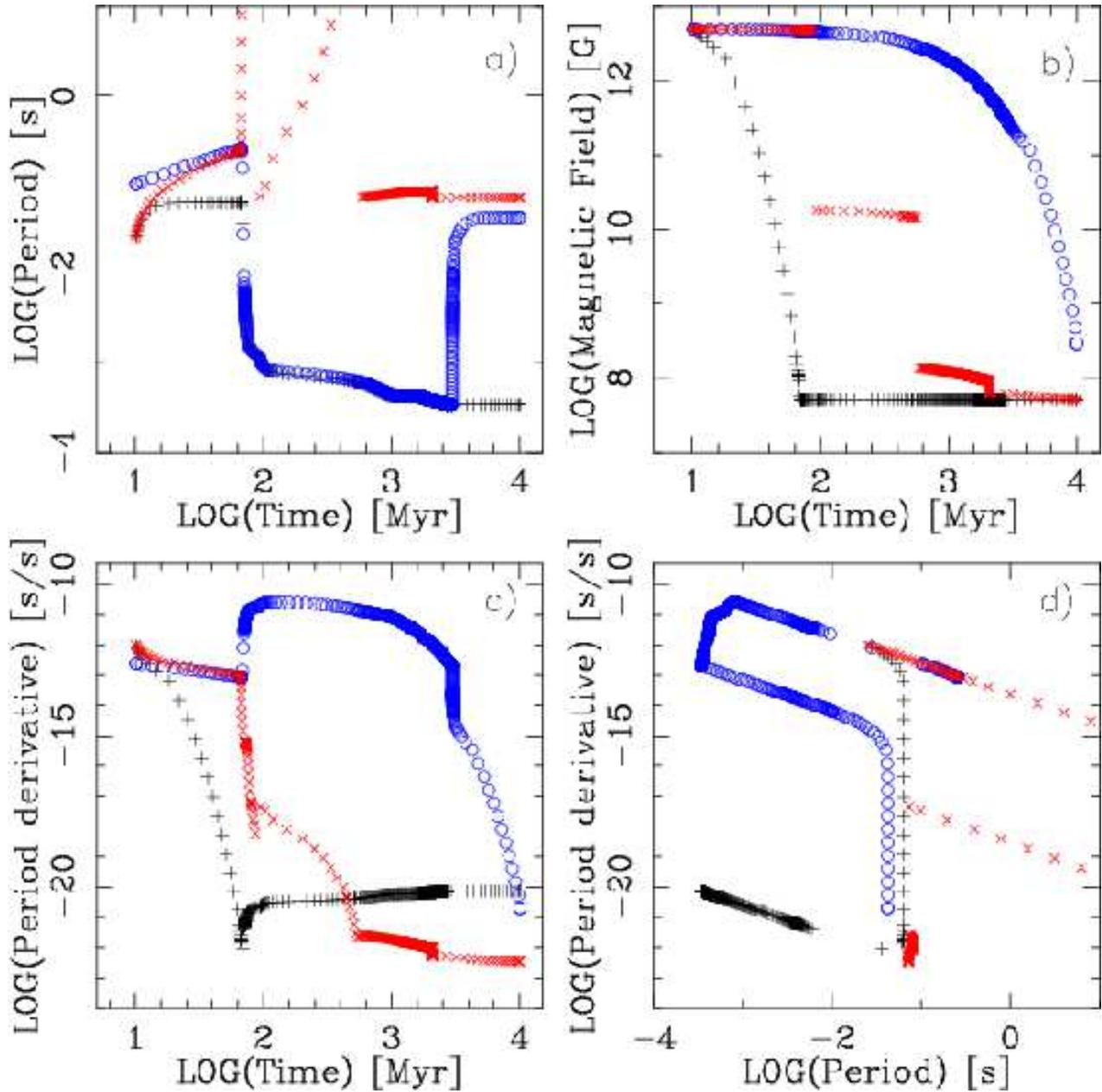}
  \caption{
  The evolution of pulsar particulars, $P$, $B_{\rm s}$ and $\dot{P}$
  for differing evolutionary assumptions, $\tau_{\rm B}$ (plus), 
  $k$ (circles) and $propeller$ (cross).
  \label{f:X2}}
\end{figure*}

\subsection{A binary MSP}

We now consider the same NS/pulsar as before but place 
it within a binary system.
The initially $20~M_\odot$ primary star (NS progenitor) 
has a $5.5~M_\odot$ companion in a circular 
$P_{\rm orb} = 32$ day ($a = 125~R_\odot$) orbit.
Initially the stars are assumed to reside on the ZAMS.
Due to the greater mass of the primary star it evolves 
more quickly than its secondary companion, leaving the 
MS at the same time as its isolated counterpart of the 
previous example (as compared to the secondary, who 
leaves the MS after 540 Myr).
The primary evolved to fill its Roche-lobe at a system 
time of $8.8~\rm{Myr}$.
At this point it is in the HG phase of evolution and has lost 
$\sim 1~M_\odot$ in a stellar wind -- the secondary accreted 
$0.285~M_\odot$ of this and was spun up accordingly.
At the onset of RLOF mass transfer occured on a dynamical 
timescale and lead to formation of a common-envelope.
During the common envelope phase the MS secondary and the 
primary core spiral in toward each other with frictional 
heating expelling the CE.
This phase is halted when the entire envelope is 
removed ($\sim 14~M_\odot$) and the two stars are just $8~R_\odot$ apart.
At this stage the MS secondary star is extremely close to 
over-flowing its Roche-lobe radius, however, the primary 
star -- now being a naked helium star -- continues to release mass 
in the form of a wind (which we assume has the form given in 
Kiel \& Hurley 2006) and due to conservation of angular momentum
begins to increase the distance between the stars.
The secondary star accretes a small fraction of this wind material 
and grows to a mass of $5.815~M_\odot$.
Then, after just $10$ Myr the primary star goes SN ejecting 
a further $3~M_\odot$ instantly from the system to become a 
$1.7~M_\odot$ NS. 
An eccentricity of $0.4$ is induced into the orbit and the 
separation is now increased to $15~R_\odot$.
By the time the secondary star evolves enough to initiate steady 
mass transfer, at a system time of $67~\rm{Myr}$, the orbit has been 
circularised by tidal forces acting on the secondary stars' envelope.
The accretion phase lasts for $2800~\rm{Myr}$.
Within this time the secondary star evolves from the MS to the HG and 
then onto the first giant branch.
During accretion the secondary loses $5.6~M_\odot$, 
the NS accretes $1.1~M_\odot$ and the separation 
increases out to $32~R_\odot$.
Beyond this the system does not change greatly.
After a short time ($\sim 30~\rm{Myr}$) the secondary evolves 
to become a $0.2~M_\odot$ helium WD (HeWD) -- being stripped 
of its envelope the WD does not develop any deep mixing to form 
a higher metalicity content in its envelope (e.g. such as 
carbon-oxygen or oxygen-neon-magnesium WDs).
At the end of the simulation we are left with a 
$2.885~M_\odot$ NS with a $0.2~M_\odot$ WD
companion in an $\sim 10~\rm{day}$ orbit.
This is typical of the parameters of observed pulsar-HeWD binaries
(van Kerkwijk et al. 2004).
We have assumed a maximum NS mass of $3~M_\odot$, however this mass 
limit is not well constrained.
If we were to assume a maximum NS mass of only $2~M_\odot$ 
the system would end as a WD orbiting a BH.

In terms of pulsar evolution, the magnetic field versus period 
parameter space covered by the NS is shown in figures~\ref{f:X1} and 
\ref{f:X2} for a variety of parameter value changes.
To begin we look at how simply changing the initial period and
magnetic field changes the NS particulars over time.
We direct the reader to Figure~\ref{f:X1} which shows the 
clear differences in the binary pulsar life if it starts its life with 
$P_0 = 2.5\times10^{-2}~\rm{s}$, $B_{\rm s0} = 5\times10^{12}~\rm{G}$ and 
therefore $\dot{P}_0 = 9\times10^{-13}~\rm{s/s}$ (pluses) or if it starts 
its life with $P_0 = 9.3\times10^{-2}~\rm{s}$, $B_{\rm s0} = 1.4\times10^
{12}~\rm{G}$ and $\dot{P}_0 = 2\times10^{-14}~\rm{s/s}$ (circles).
There is a marked difference between the spin evolution of the 
pulsars -- the evolutionary pathways depend greatly on the strength
of the pulsars magnetic field.
It is obvious from these curves where accretion began and beyond
this the pulsar evolution is similar for the two stars.
Comparison can also be made with the isolated pulsar evolution
given in the previous example (lower dots), initially these two
pulsars evolve similarly until accretion occurs.
Another isolated pulsar is shown (higher dots), with 
$P_0 = 3.4\times10^{-2}~\rm{s}$, $B_{\rm s0} = 4.1\times10^{13}~\rm{G}$ and 
$\dot{P}_0 = 4.9\times10^{-11}~\rm{s/s}$.
This isolated pulsar spins down much faster than the other three 
pulsars (see Figure~\ref{f:X1}a) and shows the effect a greater 
magnetic field has on the spin evolution of a pulsar.
Once accretion occurs the spin and magnetic field evolution
for both binary pulsars changed dramatically.
The NSs are eventually spun-up and both possess sub-millisecond 
spin periods, while the magnetic fields decay rapidly and both
reach their lowest limits ($5\times10^7~\rm{G}$) in 
$\sim 30$ years.
The $\dot{P}$ and $P$ parameter space covered by the three 
pulsars is given in Figure~\ref{f:X1}d.

At this stage we have not considered allowing the NS to receive a
velocity kick during the SN event.
If we allow a large, randomly orientated velocity kick of 
$\sim 190~\rm{km/s}$ to occur during the SN then the system
is disrupted.
However, if a smaller velocity kick is given, say, $\sim 50~\rm{km/s}$
not only does the system survive but it passes through a complex
set of mass transfer phases including two RLOF
phases and a handful of symbiotic phases.
The first RLOF mass transfer phase, which
began at a system time of $75~\rm{Myr}$, caused the NS to be spun
up to $0.02~\rm{s}$, accrete $0.04~M_\odot$ and lasted 
for $0.2~\rm{Myr}$.
The secondary star lost $4.8~M_\odot$ of mass and the separation
reduced to $5.8~R_\odot$.
During this phase the magnetic field decayed to its bottom value 
during the first Roche-lobe phase and the pulsar crosses any 
assumed death line and can not be observed beyond this evolutionary point.

\subsection{Binary evolution variations}

The examples of binary MSP formation above assumed evolutionary parameters
of $\tau_{\rm B} = 1000~$Myr, $k = 10000$ and no propeller evolution.
We now reconsider the example with 
$P_0 = 2.5\times10^{-2}~\rm{s}$ and $B_{\rm s0} = 5\times10^{12}~\rm{G}$ 
and vary these assumed parameters one at a time.
The variations are (see Figure~\ref{f:X2}):
\begin{itemize}
\item Assume $\tau_{\rm B} = 5~\rm{Myr}$ 
(plus symbols within Figure~\ref{f:X2});
\item Decreasing $k$ to $0$ (circles);
\item Allow propeller evolution to occur (crosses);
\end{itemize}

As expected a lower value of $\tau_{\rm B}$ forces the magnetic
field to decay rapidly while causing little spin-down of the pulsar
spin period to occur.
The binary evolution is not noticeably influenced due 
to this evolutionary change.

Decreasing $k$ affects the rate of magnetic field decay during accretion
(see equation~\ref{e:bacc})
as depicted clearly in Figure~\ref{f:X2}b.
This parameter change seems one we would definitely regard with
skepticism because it populates entirely the wrong region of
the $P\dot{P}$ parameter space, in terms of what observation suggest.
The binary evolution, again, does not change significantly.

Generally it is thought that if there is going to be a propeller 
phase it will occur at the beginning of the accretion phase 
(Possenti et al. 1998), when the pulsar is rotating rapidly 
enough and the magnetic field is sufficiently large enough.
This is what occurs here, however, unlike the propeller phase considered
by Possenti et al. (1998), our method causes it to
initiate multiple times during this evolutionary example.
During the first propeller phase the NS is spun-down, reaching
a spin period of $5\times10^4$ s before the phase ends and accretion
onto the NS begins.
Although the RLOF phase lasts for $\sim 2000$ Myr -- 
shorter than without a propeller phase -- the accretion of 
mass onto the slowly rotating NS occurs
extremely rapidly and unfortunately is only calculated 
by the \textsc{bse} over a single time-step.
This is the reason for the large ammount of magnetic field
decay in one step.
Modifying the time-stepping around this region does not
affect the evolution greatly, however, it does slow the
speed at which we may evolve many systems.
The first propeller phase although very efficient 
in braking the NS spin was short lived, lasting only 
$\sim 2.3 $ Myr.
While beyond the first mass transfer phase the system 
moved once again into propeller evolution, this time 
lasting for $\sim 550$ Myr and spinning the NS down to a
$\sim 1000$ s spin period.
Once again, at a system time of $\sim 580$ Myr the NS 
accretes material and is spun in one time-step up to 
$\sim 0.04$ s spin period.
The third propeller phase, initiated directly after the 
accretion phase, continues until the end of the RLOF mass 
transfer phase and manages to spin the NS to a period $\sim 0.08$ s.
At the end of the RLOF phase the companion, now on the
first giant branch and $40~R_\odot$ away from the NS, loses 
$\sim 0.001~M_\odot$ of matter the majority of which is 
collected by the NS.
This spins up the pulsar to $0.07$ s.
We end with a $0.3~M_\odot$ HeWD, slightly greater than 
that created without the propeller included.
While the NS is $1.78~M_\odot$, slightly smaller than when 
we do not assume the propeller phase to occur.

From the simple analysis above we can see how effective the 
propeller phase is at braking the pulsar spin.
Although the propeller phase arrests the majority of spin-up, 
and in fact obstructs the formation of a MSP 
(that would otherwise have formed), it is an important 
evolutionary assumption to test on populations of NSs.
We have also shown that both the accretion-induced 
field decay parameter $k$ and the standard magnetic field time-scale
parameter $\tau_{\rm B}$ are important parameters to regard when 
modeling the entire pulsar population.
We next consider populations of pulsars and how parameter changes
affect the morphology of these populations.

\section{Results and Analysis}

\subsection{Method}
\label{s:method}

A binary system can be described by three initial parameters: 
primary mass, $M$, secondary mass, $m$, and separation, 
$a$ (or period, $P_{\rm orb}$). 
A fourth parameter, eccentricity, may also vary, however, for simplicity 
we assume initialy circular orbits.
For the range of primary mass we choose a minimum of 
$5~M_\odot$ and a maximum of $80~M_\odot$. 
To first order this covers the range of initial masses that will evolve
to become NSs via standard stellar evolution.
Stars below $\sim 10~M_\odot$ will generally become WDs and stars above
$\sim 20~M_\odot$ will end up as BHs.
However, binary interactions tend to blur these stellar mass limits.
As sugested in HTP02 the production of stars with 
$M > 80~M_\odot$ is rare when using any reasonable IMF,
hence the choice of upper mass limit. 
The lower limit is one that should allow the production of all interesting 
NS systems to be formed, that is, all systems which have interaction between
the NS and its companion.
This includes stars with initial masses below $10~M_\odot$ that accrete mass
and form a NS instead of a WD.
Expanding the upper limit accounts for stars with $M > 20~M_\odot$ that
lose mass and become a NS instead of a BH.
The secondary mass range is from $0.1~M_\odot$ to $40~M_\odot$.
The lower limit here is approximately the minimum mass of a 
star that will ignite hydrogen-burning on the MS.
These mass ranges will allow the production of the majority of NS binaries
and enable us to investigate complete coverage of the $P\dot{P}$ diagram.
For the initial orbital period we consider a range between $1$ and 
$30000$ days.
The birth age is randomly selected between $0$ and $10\,$Gyr where the 
latter reflects an approximate age of the Galaxy.

Instead of calculating birth rates and numbers of systems of interest, 
our focus, here, is on comparison with the observed pulsar $P$ vs $\dot{P}$
diagram and what this can teach us about uncertain parameters
in pulsar evolution.
These include parameters such as the magnetic field decay time 
constants in both the isolated and accreting regimes along with
the effectiveness of modelling propeller systems to spin-down 
pulsars to observed periods.
We also aim to test how different evolutionary parameter changes affect 
the relative populations of binary and isolated pulsars.

In distributing the primary masses we use the 
power law IMF given by Kroupa, Tout \& Gilmore (1993),

\begin{eqnarray}
\Phi \left( M \right) = \left\{ \begin{array}{ll}
         0.035M^{-1.3} & \textrm{if $0.08 \le M < 0.5$,}\\
         0.019M^{-2.2} & \textrm{if $0.5 \le M \le 1.0$,} \\
         0.019M^{-2.7} & \textrm{if $1.0 \le M < \infty$.}
    \end{array} \right.
\end{eqnarray}
This is the probability that a star has mass $M$ [$M_\odot$]. 
The distribution of secondary star masses is not as well constrained
observationally.
We assume that the initial secondary and primary masses are 
closely related via a uniform distribution of the mass-ratio
(Portegies Zwart, Verbunt \& Ergma 1997; HTP02),
\begin{equation}
\varphi \left( m \right) = \frac{1}{M},
\end{equation} 
where the mass-ratio ranges from $0.1/M$ to $1.0$
The initial orbital period distribution we take to be  
flat in $\log\left(P_{\rm orb}\right)$.
Completely sampling these distributions allows us the full range 
and distribution of initial parameters, so we are modelling
the natural selection of stellar systems as suggested by 
observations.

The results of six simulations (see Table~\ref{t:table1}) in the 
form of $P\dot{P}$ plots are given below
(Sections~\ref{s:Mod1} to~\ref{s:windaccres}). 
These are Models A through F.
We begin by simulating a rather naive stellar population (Model A) to 
which we build pulsar evolutionary phases and assumptions upon in 
increasingly greater complexity until we feel the results best 
represent the required pulsar $P\dot{P}$ parameter space as compared 
to observations.
We note from the ouset that we do not attempt to model the
slow rotating high magnetic field pulsars in the upper-right region 
of the $P\dot{P}$ parameter space.
As discussed in Section~\ref{s:disc} we leave these potential 
magnetars for future work.
Modifying the assumed initial NS spin period and magnetic field 
(see Section~\ref{s:updatedist}) quantifies the importance of choices 
made in this area.
For the most part the initial period and magnetic field ranges 
(given in Table~\ref{t:table1}) are selected so as to allow good 
coverage of the $P\dot{P}$ parameter space.
At this stage we do not suggest that these ranges are completely realistic
but we do wish to show that it is possible to produce a complete coverage of
the pulsar $P\dot{P}$ parameter space with them.
In Section~\ref{s:pref} we take Model F and look at some further 
variations such as inclusion of electron 
capture supernovae and beaming effects.
This includes an examination of the pulsar primordial parameter 
space (Section~\ref{s:Primordial}) and how changing evolutionary assumptions
modifies the pulsar birth properties.

Unless otherwise stated all models have the binary and stellar 
evolutionary parameters of: solar metalicity $Z = 0.02$; a
maximum possible NS mass of $3~M_\odot$ and $\alpha_{\rm CE} = 3$.
SN kicks are given to NSs, while to keep the required number of 
models down to a minimum we only use the curvature radiation 
death line model with $\bar{f}^{\rm min}_{\rm prim} = 0.15$.
Although we know the CR death line to be insufficient in terms
of modelling recycled pulsars it is adequate for our purposes here
(for more see Section~\ref{s:disc}).
For each model we evolve $10^7$ primordial systems, all 
of which are initially binary systems.
Integrating over the Kroupa, Tout \& Gilmore IMF with the mass limits set
above, modeling $10^7$ primordial systems is roughly representative of
$2\times10^9$ systems in the Galaxy if the full mass range is allowed.
One way to think of this is that we are assuming a stellar
birth rate of $10^{-3}$ systems/year, assuming the age of the Galaxy
is $10$ Gyr.
In mass, we are only evolving $\sim 0.1\%$ of the Galaxy assuming it
contains $\sim 1 \times 10^{11}~M_\odot$.
This is something that in the future we will wish to increase.

\begin{table*}
 \centering
 \scriptsize
 \begin{minipage}{140mm}
  \caption{Characteristics of the main set of models used in this work. 
    The first column gives an identifying letter for each model. 
    Within the second column we indicate the value of the magnetic 
    field decay time-scale which is followed by the choice of accretion 
    induced magnetic field decay parameter. 
    The fourth and fifth columns indicate whether or not propeller 
    evolution is considered and whether a link between SN 
    strength and initial magnetic field and pulsar period is assumed. 
    The next four columns indicate the maximum and minimum values 
    for the initial distributions of pulsar spin period and magnetic 
    field.
    We then modify the ammount of angular moment accreted by a NS when
    the companion is lossing mass in a wind, this is governed by the 
    parameter $\Xi_{\rm wind}$.
    The penultimate column indicates if the electron capture SN kick 
    distribution is considered and the final column indicates whether
    beaming and accretion selection effects are considered.
  \label{t:table1}}
  \begin{tabular}{crrrrrrrrrrr}
  \hline
 Model & $\tau_{\rm B}$ [Myr] & $k$ & Propeller & SN link & $P_{\rm 0min} \rm{[s]}$\footnote{
     For the SN link models this value is the 
     average initial spin period $P_{\rm 0av}$, not the minimum.} & 
     $P_{\rm 0max} \rm{[s]}$ & $B_{\rm 0min} \rm{[G]}$ & 
     $B_{\rm 0max} \rm{[G]}$\footnote{For the SN link models this value is 
     the average initial magnetic field $B_{\rm 0av}$, not the maximum.} & 
     $\Xi_{\rm wind}$ & EC SN & Beaming \\
\hline 
 A & $1000$ & $0$ & No & No & $0.01$ & $0.1$ & $1\times10^{12}$ & $3\times10^{13}$ & 1 & No & No \\
 B & $1000$ & $10000$ & No & No & $0.01$ & $0.1$ & $1\times10^{12}$ & $3\times10^{13}$ & 1 & No & No \\
 C & $1000$ & $10000$ & Yes & No & $0.01$ & $0.1$ & $1\times10^{12}$ & $3\times10^{13}$ & 1 & No & No \\
 D & $5$ & $10000$ & Yes & No & $0.01$ & $0.1$ & $1\times10^{12}$ & $3\times10^{13}$ & 1 & No & No \\
 E & $2000$ & $3000$ & Yes & Yes & $0.02$ & $0.16$ & $5\times10^{11}$ & $4\times10^{12}$ & 1 & No & No \\
 F & $2000$ & $3000$ & Yes & Yes & $0.02$ & $0.16$ & $5\times10^{11}$ & $4\times10^{12}$ & variable & No & No \\
 Fb & $2000$ & $3000$ & Yes & Yes & $0.02$ & $0.16$ & $5\times10^{11}$ & $4\times10^{12}$ & variable & Yes & No \\
 Fc & $2000$ & $3000$ & Yes & Yes & $0.02$ & $0.16$ & $5\times10^{11}$ & $4\times10^{12}$ & variable & Yes & Yes \\
 Fd & $2000$ & $3000$ & No & Yes & $0.02$ & $0.16$ & $5\times10^{11}$ & $4\times10^{12}$ & variable & No & No \\
\hline
\end{tabular}
\end{minipage}
\end{table*}

\subsection{A starting model}
\label{s:Mod1}

\begin{figure}
  \includegraphics[width=84mm]{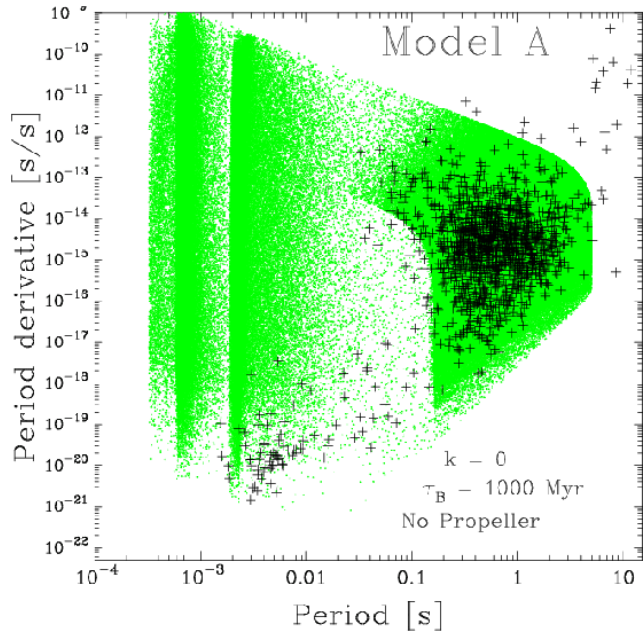}
  \caption{
The $P\dot{P}$ diagram for Model A. 
The relevant parameter values are given in Table~\ref{t:table1} 
and on the figure.
Plus symbols represent the observed pulsars, while grey dots represent
each pulsar the simulation produces.
  \label{f:res1}}
\end{figure}

We start by demonstrating the results of pulsar population
synthesis when no field decay owing to the accretion of
material onto a NS is allowed.
This is done by setting $k = 0$ in Equation~\ref{e:bacc}.
The result is Model A for which the pulsar evolutionary parameters 
are given in Table~\ref{t:table1} and the $P\dot{P}$ distribution
is shown in Figure~\ref{f:res1}.
Clearly from Figure~\ref{f:res1} there are a great number of 
rapidly rotating pulsars with large spin period derivatives.
This is not observed which means that this model fails to 
decay the magnetic field quickly enough during accretion.
Therefore, Model A seems physically unrealistic.
One may be skeptical of this result based on one model, however,
similar $P\dot{P}$ distributions occur even if we vary additional 
parameters such as $\Xi_{\rm wind}$ (to lower the angular 
momentum transfered during accretion) and/or using 
lower values of $\tau_{\rm B}$.
It is interesting to observe the different levels of pulsar 
densities in the spun-up region of Figure~\ref{f:res1} -- 
there is almost a wave-front effect towards MSP periods.
The pulsars in these regions have differing companion types and 
orbital configurations indicating different mass transfer histories.
Here the fastest rotating pulsars generally have MS companions 
and circular orbits.
In these cases mass transfer occurs via RLOF.
For the pulsars located in the spin-up region truncated at 
$P\sim 0.002~$s companion types are mainly 
giant and white dwarfs and include many cases of eccentric orbits.
This indicates mass transfer via a stellar wind from a giant star 
in a binary with a relatively long orbital period.
Systems which have a shorter orbital period and initiate RLOF from
a MS companion can accrete a greater ammount of material and hence 
are spun-up further to left in the $P\dot{P}$ diagram.

\subsection{Magnetic field decay during accretion}
\label{s:accres}
\begin{figure}
  \includegraphics[width=84mm]{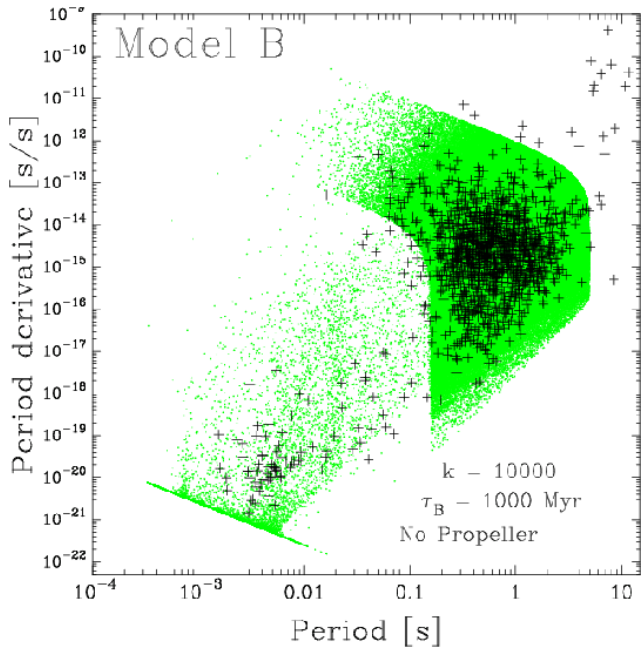}
  \caption{
The $P\dot{P}$ diagram for Model B. 
The relevant parameter values are the same as Model A except for
the accretion induced field decay parameter which is now $k = 10000$.
Plus symbols represent the observed pulsars, while grey dots represent
each pulsar the simulation produces.
  \label{f:res1dash}}
\end{figure}

Model A shows that some form of accretion induced magnetic
field decay is required to suitably populate the 
recycled region of the $P\dot{P}$ diagram.
A first attempt at modelling this physics is completed in
Model B (see Table~\ref{t:table1}), where we $k = 10000$ 
in Equation~\ref{e:bacc}.
The resultant $P\dot{P}$ parameter space is shown in 
Figure~\ref{f:res1dash}.
This simulation gives good coverage of the observed 
pulsar regions -- the standard and recycled islands.
However, there is an over-production of systems linking the two
pulsar $P\dot{P}$ islands.
Many of these pulsars are paired with early type stellar
companions (pre-degenerate stars) with strong winds that may
disrupt the radio beam (see Section~\ref{s:Beam}).
A number of these systems are also beyond what is considered the 
general `spin-up' line.
The spin-up line is a theoretical construct which suggests that
an accreting pulsar will eventually reach some equilibrium spin 
period and therefore end pulsar spin-up during the accretion
phase.
The spin-up line, however, is a very uncertain model and the
majority of assumptions that go into it are highly variable 
(see Arzoumanian, Cordes \& Wasserman 1999 and references therein
for details), and the line is sensitive to changes for many parameter 
values.
We do not directly impose the spin-up line in our models but instead
aim to have its effect replicated by the physical processes that
make up our binary/pulsar evolution algorithm.
As such we would clearly want to limit the production of medium-low 
period high-$\dot{P}$ pulsars that appear in Model B.
This is a motivator for testing the affect of including propeller physics.
One may also notice the line of MSPs at the bottom right of 
Figure~\ref{f:res1}.
This line appears in all the $P\dot{P}$ figures shown here and
is caused by the assumed limit of magnetic field decay (Section~\ref{s:Bbot}).

\subsection{Propeller evolution}
\label{s:Mod3}
\begin{figure}
  \includegraphics[width=84mm]{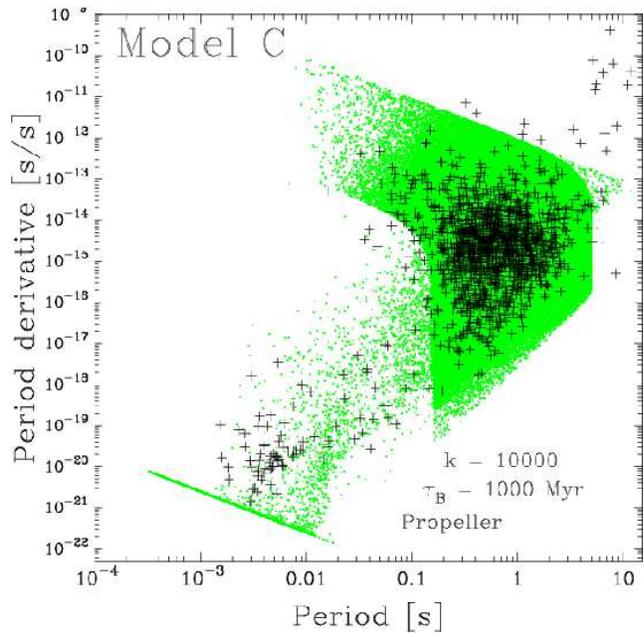}
  \caption{
The $P\dot{P}$ parameter space for Model C.
Plus symbols represent the observed pulsars, while grey dots represent
each pulsar the simulation produces.
  \label{f:res3}}
\end{figure}

Propeller evolution (Section~\ref{s:Paccev}) is permitted in Model C, 
which is otherwise the same as Model B.
The results are shown in Figure~\ref{f:res3} which exhibits three 
main features that make modelling the propeller evolution beneficial.
The first is the slight bump from the standard pulsar island 
in slower spin period pulsars ($P > 5~\rm{s}$).
This slight bump is directly caused by the propeller phase,
and allows us to model those pulsars who reside next to the 
standard pulsar island.
Although, we note that those pulsars could also be modelled if we assume that
the magnetic field does not decay other than during accretion.
The second feature of interest is the cut off of millisecond pulsars 
with period derivatives between $10^{-21}$ and $10^{-18}$, as compared
to Model B.
Although this cut-off occurs too soon (in terms of pulsar period) it
does provide a method for producing this observed limit in period.
The third feature is lack of pulsars with period derivatives 
between $10^{-16}$ and $10^{-13}$ and spin periods of around $0.01$.
Therefore, when considering $P\dot{P}$ parameter space, we find that 
including the propeller phases provides much better 
agreement with models that impose a spin-up line directly to restrict 
pulsars appearing above the observed pulsar bridge.

\subsection{Magnetic field decay timescale}
\label{s:Mod2}
\begin{figure}
  \includegraphics[width=84mm]{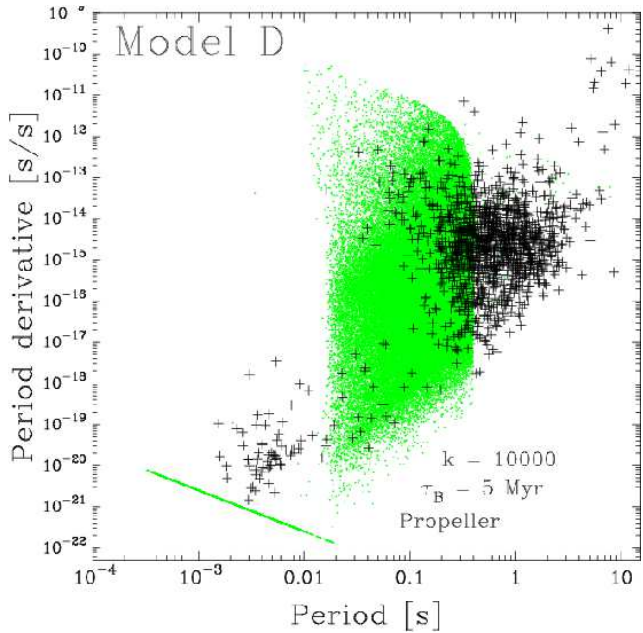}
  \caption{
The pulsar $P\dot{P}$ parameter space for observed and 
those simulated pulsar produced by Model D.
This facilitates comparison between a large and small 
$\tau_{\rm B}$ when compared to Figure~\ref{f:res3}.
Plus symbols represent the observed pulsars, while grey dots represent
each pulsar the simulation produces.
  \label{f:res4}}
\end{figure}

Model D differs from the evolutionary assumptions in Model C 
by assuming $\tau_{\rm B} = 5~\rm{Myr}$ (reduced from $1000~\rm{Myr}$).
This rapid field decay has the effect of shifting the standard pulsar island
to the left in the $P\dot{P}$ diagram (see Figure~\ref{f:res4}).
Even so, looking at Figure~\ref{f:res4} we see it is not possible to 
properly simulate the standard pulsar island with $\tau_{\rm B} = 5$ Myr,
unless a completely contrived initial period range is used.
Also in Figure~\ref{f:res4} it is noticable that even before consideration 
of selection effects there is a large underproduction of recycled pulsars.
The rapid field decay causes the majority of spun-up pulsars to sit on the 
lower magnetic field limit, a region of the $P\dot{P}$ parameter 
space in which any complete death line model should remove.
As such, assuming such a low value of magnetic field decay, as has been 
used in the past by Gonthier et al. (2002), for example, is unrealistic.
We find that $\tau_{\rm B}$ of 100 Myr or greater is required to 
produce a reasonable $P\dot{P}$ distribution.

\subsection{Updated initial pulsar distribution: SN link}
\label{s:updatedist}

\begin{figure}
  \includegraphics[width=84mm]{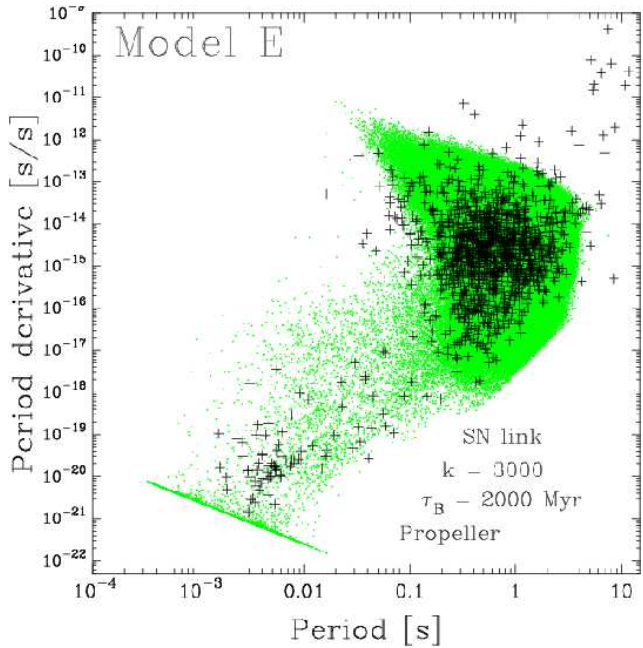}
  \caption{
The pulsar $P\dot{P}$ parameter space 
for Model E.
Plus symbols represent the observed pulsars, while grey dots represent
each pulsar the simulation produces.
  \label{f:res7}}
\end{figure}

With the propeller phase producing what seems to be a more 
realistic population in Model C than that which occurred in 
Model B, while also constraining $\tau_{\rm B} \geq 100~$Myr 
in Model D, we now conduct a trial to see what effect linking 
the SN velocity kick magnitude with the initial period and 
magnetic field (see Section~\ref{s:InitPparams}) has on the pulsar 
distribution in $P\dot{P}$.
To do this we make use of equations~\ref{e:Pinit2} and \ref{e:Binit1}, 
assuming $n_{\rm p}$ and $n_{\rm b}$ are unity.
The corresponding values of $P_{\rm 0av}$ and $B_{\rm 0av}$ are given in 
Table~\ref{t:table1} (see Model E).
Note that the distribution of initial pulsar period and magnetic
field is modified (as compared to Models A through C)
to allow better coverage of the standard pulsar island in the 
observed $P\dot{P}$ space.
The value of the lower initial magnetic field limit, $B_{\rm s0min}$, 
was used due to suggestions that there are NSs which are born 
weakly magnetised ($\sim 5\times10^{11}$ G; Gotthelf \& Halpern 2007).
We also allow a greater coverage of the slower rotating pulsars
by assuming $\tau_{\rm B} = 2000~$Myr.
The previous value of $\tau_{\rm B} = 1000~$Myr under populated the 
slower pulsars when used in conjunction with this model.
Conversly if the value of $\tau_{\rm B}$ is greater than $2000~$Myr 
the lower region of the standard pulsar island is increasingly 
truncated -- the magnetic field decays too slowly to sufficiently
decrease the period derivative.
These changes result in Model E and its $P\dot{P}$ diagram is shown in 
Figure~\ref{f:res7}.
There are now striking similarities between the observed and simulated
$P\dot{P}$ parameter spaces, particularly at the standard pulsar island.
Three main features in Figure~\ref{f:res7} of worthy note are 
as follows:
(1) the production of a wall of young pulsars, which indicates 
that those pulsars born with strong magnetic fields also 
initially spin at a faster rate to those born with lesser magnetic fields 
(which is what we expect from Equations~\ref{e:Pinit2} and ~\ref{e:Binit1});
(2) the only method for pulsars to have periods of $\sim 4$ seconds or 
slower, in this model, is to pass through a propeller phase;
and (3) the observed group of high $\dot{P}$ ($\sim 10^{-14} - 
10^{-12}~\rm{[s/s]}$) but moderately rapid rotating pulsars 
($\sim 0.01 - 0.2~\rm{[s]}$) are reasonably well modelled. 
It should be pointed out here that within Model E there is
a different reason why the $P\dot{P}$ area described in point three 
is populated as compared to the explanation given for Figure~\ref{f:BvsP}.
Within Model E these systems arise due to mass transfer, while the
assumed formation mechanism of these pulsars (for Figure~\ref{f:BvsP})
was that these where newly formed NSs and born there.

In terms of the shape of the $P\dot{P}$ parameter space, we are able
to simulate some regions of the observations rather accurately.
To begin we only slightly over-estimate the cut-off of pulsars at
small spin periods.
This cut-off arises from a combination of the assumption of $k$ and 
the propeller evolutionary phase.
The death line does a satisfactory job at limiting the semi-recycled 
pulsars however it fails in limiting both the slower rotating pulsars
and the fully recycled pulsars.
This failure is not unexpected because the curvature radiation death line
is not a completely realistic model when considering the full pulsar 
population (see Section~\ref{s:disc} for futher discussion on this point).

What is not noticable in Figure~\ref{f:res7} is the number of pulsars
that evolve through a Thorne-\.{Z}ytkow object phase. 
As explained earlier a T\.{Z}O is a remnant core (NS or BH) 
surrounded by an envelope formed from the non-compact 
star with which it has merged.
In \textsc{bse} these objects are treated as unstable so that
the envelope is ejected instantaneously.
However, we are left with the question of what to do with the
pulsar parameters.
For now we keep this evolutionary phase consistent with our 
CE phase were we assume that the spin period and magnetic field 
are re-set by this process.
This assumption is prefered at this stage because if we assume
the pulsar is left unchanged we find many isolated pulsars rotating
near their assumed break-up spin velocities ($P \sim 4\times10^{-4}$ s)
with a large range of spin period derivatives ($\dot{P} < 10^{-15} $).
We note here that our assumption regarding these objects 
is purposefully simplistic to reflect our lack of knowledge about what
occurs during the formation and evolution of a T\.{Z}O but
other possibilities are discussed in 
Sections~\ref{s:Beam} and~\ref{s:disc}.

Once again, there is an over production of pulsars 
above the observed pulsars which link the two pulsar islands.
In the next section we again increase the detail of our model 
in an attempt to constrain this $P\dot{P}$ region.

\subsection{Testing wind angular momentum accretion, the $\Xi_{\rm wind}$ parameter}
\label{s:windaccres}
\begin{figure}
  \includegraphics[width=84mm]{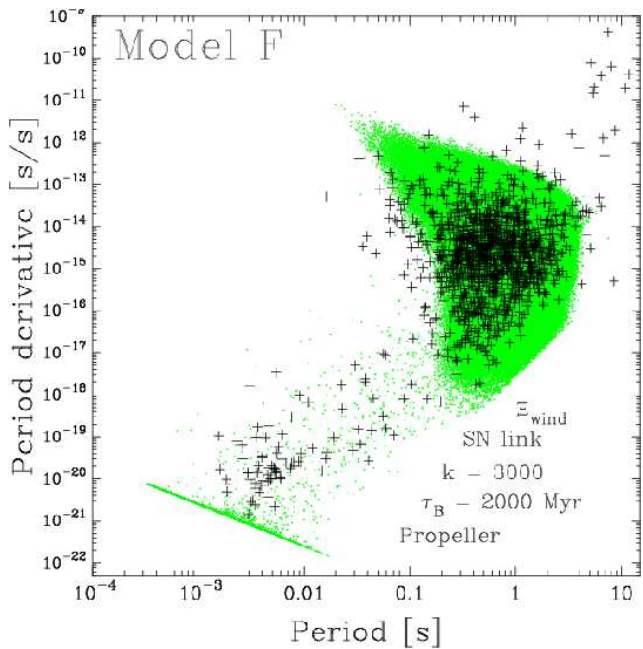}
  \caption{
The pulsar $P\dot{P}$ parameter space 
for Model F.
Plus symbols represent the observed pulsars, while grey dots represent
each pulsar the simulation produces.
  \label{f:res8}}
\end{figure}
\begin{figure}
  \includegraphics[width=84mm]{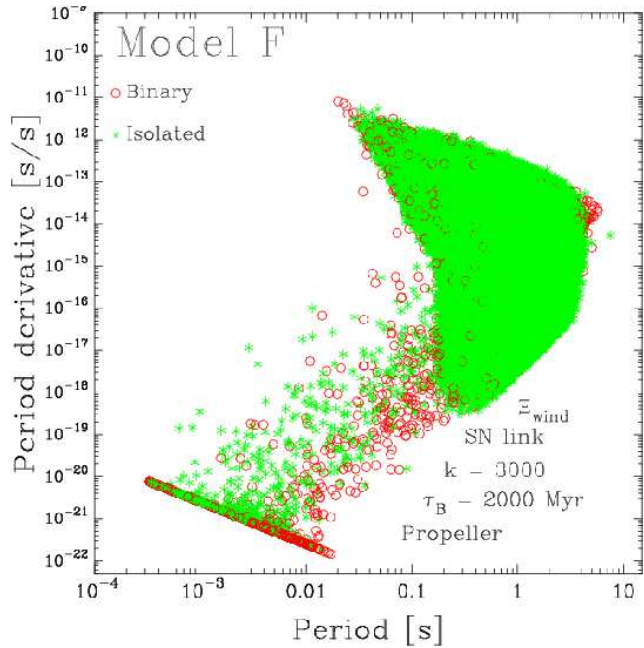}
  \caption{
The binary (circles) and isolated (stars) pulsar $P\dot{P}$ parameter space 
for Model F.
  \label{f:res8_1}}
\end{figure}

The effect of modifying $\Xi_{\rm wind}$ (Model F) is shown in 
Figure~\ref{f:res8} and is dramatic when compared with Figure~\ref{f:res7}.
This modification truncates the width of the pulsar bridge 
which was the aim of this model.
The pulsar bridge that stretches between the two pulsar islands
compares rather accurately to the observations (in terms of $P\dot{P}$ 
area covered).
This is our prefered model and we now perform an in depth analysis 
of this model, including further parameter changes (see 
Table~\ref{t:table1}).

\subsection{Model F: analysis and further additions}
\label{s:pref}

Before we embark on our further examination of Model F we must
remind the reader that -- at this stage -- our parameter space
analysis is incomplete.
Therefore, although we provide evidence for the formation
of such systems as: isolated MSPs, eccentric double neutron stars and 
possible existence of Galactic disk eccentric MSP-BH binaries;
we must stress that it is possible for many realistic model parameter 
changes to modify the resultant formation probability of these systems.
However, we provide this section as an example of typical 
binary, stellar and pulsar features that our future complete pulsar 
population synthesis package will be able to constrain when one 
is able to compare theoretical results directly with observations.
This section also provides evidence of how some evolutionary 
assumptions further modify the final pulsar population.

\begin{figure}
  \includegraphics[width=84mm]{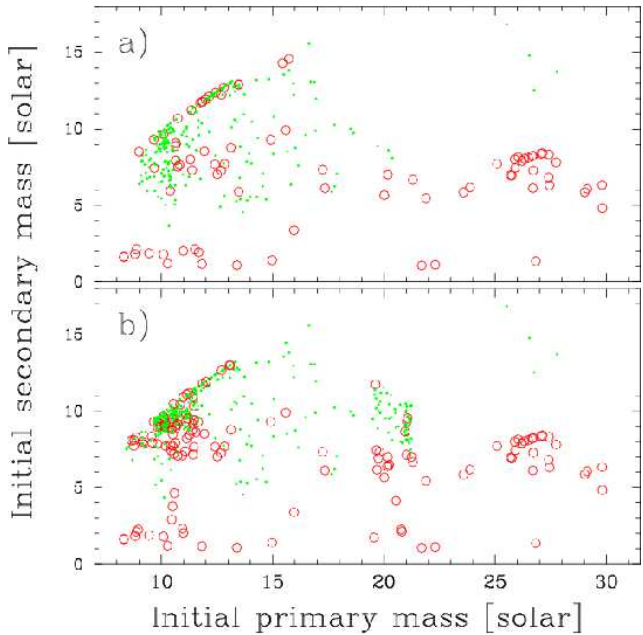}
  \caption{
The binary (circles) and isolated (dots) MSP initial mass 
parameter space for Model F in a and Model Fb in b.
All systems here fulfill the
$6\times10^7 \le B_{\rm s} \le 1\times 10^9$ G and $P < 0.02$ criterion. 
  \label{f:res8_2}}
\end{figure}

\label{s:mspratio}
Figure~\ref{f:res8_1} depicts the isolated and binary pulsar systems 
in the $P\dot{P}$ parameter space for Model F.
Focusing on MSPs, the relative number of binary to 
isolated MSPs is $0.49$.
Note that we are restricting our comparison to those pulsars 
that have a magnetic field greater than $6\times10^7$ G and less 
than $1\times10^9$ G
(model pulsar period region is  
$P < 0.02$; see Figure~\ref{f:BvsP} for observational comparison).
In doing this we are ignoring those pulsars that sit on the 
bottom magnetic field limit.
To examine the MSP population of Model F in further detail 
we show the initial primary and secondary mass 
distributions of the isolated and binary MSPs in 
Figure~\ref{f:res8_2}a and see that these differ.
Notably the average initial mass of the secondary stars is 
typically greater for binaries that go on to produce isolated MSPs.
What this is showing is that the isolated MSPs are passing 
through two SN events and it is the second SN that disrupts 
the system.
This provides us with a simple method for producing isolated
MSPs rather than requiring the addition of some ad-hoc model 
where the MSP is destroying the companion.
This suggested formation mechanism of isolated MSPs is 
not restricted to Model F alone but to all models completed so far, 
in fact, due to our restriction of the accreted wind angular momentum 
we are now producing less isolated MSP systems than previously.
Therefore, we believe this is an important evolutionary phase
that should be placed under further scrutiny within the community.

A number of these disrupted MSPs arise from systems which contain 
initial secondary masses lower than the assumed upper mass limit 
for the EC SN model as shown in Figure~\ref{f:res8_2}.
This suggests that if we allow EC SNe and adopt the associated 
lesser $V_\sigma$ value (see Section~\ref{s:ecsn}) that many of
the now isolated MSPs may retain their binary companion.
This leads us to consider the effect the combined
EC SN and standard SN velocity kick distribution has on 
our model pulsar population.

\subsubsection{A new supernovae velocity kick distribution: electron capture supernovae}
\label{s:ecsnres}

We have repeated Model F with the electron capture SN kick distribution 
taken into consideration to produce Model Fb.
We find that Models F and Fb do not differ 
greatly from each other in terms of the regions of the
$P\dot{P}$ parameter space covered (hence we do not show Model Fb).
Surprisingly the binary to isolated MSP ratio has only risen to
$0.53$ in comparison to Model F.
However, Model Fb does contain a greater number of binary
MSPs.
The initial mass parameter range of MSPs that arise in Model Fb is 
shown in Figure~\ref{f:res8_2}b).
It is obvious that many more binary MSP
systems are now being born in and around the $\sim 10~M_\odot$
primary mass and $\sim 7 - 10~M_\odot$ secondary mass range
as expected.
Not so expected was the number of newly formed 
isolated MSPs that are also derived from this primary and 
secondary mass region.
It appears that many primary mass stars are forming NSs
via the electron capture method, allowing them to remain
within a binary system and receive angular momentum from
their companion which then also goes SN, disrupting the system.
In particular as Figure~\ref{f:res8_2}b) shows, there are a greater
number of isolated (and binary) MSPs being born from the $20~M_\odot$
primary mass region than in Model F (Figure~\ref{f:res8_2}a).
A thorough analysis of the ECSN model can not be completed 
at this stage as considerations of NS space velocities are required
for any complete examination of this assumption.
However, we have demonstrated some effects the 
ECSN model imparts on the pulsar population.
As such it seems likely that the EC assumption will play an important 
role in future NS population synthesis works.
Especially if comparison to observations of orbital period, 
eccentricity and space velocities are attempted.

\subsubsection{Beaming effect and accretion: a start on selection effects}
\label{s:Beam}
The source of radio emission from a pulsar occurs at the magnetic poles,
otherwise known as the polar caps.
The radio waves are well collimated and form a beam. 
This beam sweeps across the sky (as observed from the pulsar) and may be
observed if the beam crosses the observer line of sight.
Some regions of the pulsar sky are not covered by the beam -- 
the beaming effect considers the chance that a particular pulsar 
is beaming across our line of sight.
Tauris \& Manchester (1998) examined this and found the 
fraction $f(P)$ of pulsars beaming towards us in terms of pulsar spin period,
\begin{equation}
\label{e:beambag}
f\left( P \right) = 0.09\left( \log\left(P\right) - 1\right)^2 + 0.03.
\end{equation}
This equation shows that the faster the pulsar rotates the greater 
the chance that it is beamed towards us.
The basic reason for this is that the pulsar cap angle (beam size or beam 
angle) is larger for fast rotators.
A full explanation relies on details of the magnetic field and beam emission
mechanism and we suggest the reader consider Tauris \& Manchester (1998).
However, the pulsar still may not be observed if selection effects 
conspire against our observing it.
In fact there are some systems that we can rule out directly as not being 
observable in the radio regime.
These are accreting systems or those in which the companions wind
is energetic (Illarionov \& Sunyaev 1975; Jahan-Miri \& Bhattacharya 1994;
Possenti et al. 1998).
Thus, using Equation~\ref{e:beambag} we ignore those pulsar 
that are beamed away from us in combination with ignoring those 
that have a non-degenerate companion (including stars on the main sequence, 
although this may be too strict and could possibly be relaxed), 
or a pulsar which is accreting material.
We also assume that it takes time for a pulsar to be
observed (at radio wavelengths) after the end of an accretion phase 
-- here we assume it takes $1$ Myr.
Furthermore, any system which passed through a T\.{Z}O phase
is considered to be unobservable owing to the 
ejected nebulous material inhibiting radio transmission.
Model Fc results from these considerations.

Because this model only culls the number of pulsars in the $P\dot{P}$
parameter space we do not show a comparison figure.
It is reassuring that the inclusion of beaming does not alter
the good agreement found between the model and the observations
in terms of the $P\dot{P}$ diagram.
With the inclusion of beaming and the assumption that pulsars can 
not be observed for some time after accretion 
we find the binary/isolated MSP ratio is $0.38$ for Model Fc
with the majority of MSPs produced by wind accretion.

\subsubsection{Propeller physics appraisal}

Here we examine further the effect propeller physics (described in 
Section~\ref{s:Paccev}) has on the pulsar population.
Motivation for this arises from the lack of binary MSPs 
coming from the RLOF channel in Model Fc.
This arises because the propeller phase halts and then reverses 
the majority of spin-up phases (as seen in Figure~\ref{f:X2}).
In other words, many Roche-lobe overflow 
NS systems are having their spin-up phases truncated at some medium
to slow spin period and therefore are not being `observed' by \textsc{binpop}
as they reside below the death line within the $P\dot{P}$ parameter space
(at spin periods of $\sim 0.03$ or slower).
This is in stark contrast to the standard method of assumed MSP formation, 
which is thought to occur via a low-mass X-ray binary phase (see Section 1
and references therein; although see also Deloye 2007 for an interesting 
discussion on the low-mass X-ray binary-MSP connection).
This result begs the question `what would these later models (that is, 
say, Model F) produce if propeller evolution was not included?'.
To answer this we have repeated Model F with the propeller 
option turned off to create Model Fd.

In comparison to Model F we find that the $P\dot{P}$ parameter
space of Model Fd is similar, the most notable exception 
being a slight increase in the width of the pulsar bridge (in $\dot{P}$).
We certainly feel it is a plus to have the propeller option in our 
evolution algorithm, especially when one considers the claims of 
Winkler (2007), who suggests a new class of high-mass X-ray binary with
the massive companions losing mass in winds yet the NSs have spin periods 
of order $100 - 1000~$s.
Conversely, observations of X-ray binaries that contain a rapidly rotating NS, 
including the growing range of observed millisecond X-ray powered pulsars 
(Galloway 2007; Krimm et al. 2007), which we can produce in Model Fd,
confirm that care must be taken when implementing propeller evolution.
This suggests that improvements are required in how,
and to what extent, propeller evolution is modelled.
Our work presented here also corresponds to the 
claims made by Kulkarni \& Romanova (2008).
They show that unstable transfer of mass onto the accretor may occur 
-- even during stable mass transfer events.
Therefore, further investigation in this 
area is required, especially with regard to the possibility of 
transient pulsar mass accretion in X-ray binaries.

It is interesting to note that with the increased number of RLOF binary
MSPs in Model Fd that the binary/isolated ratio is $1.6$.
This compares well to the observed ratio (Huang \& Becker 2007).
However, it is too early on in our study to make conclusions based
on such a comparison and there may well be a set of parameter changes
that affect this result that we have yet to consider.

\subsubsection{Primordial parameter range}
\label{s:Primordial}

Up until now the majority of our focus has been on pulsar evolution
and how modifying the parameter values used in the assumed evolution
of pulsars affects the distribution of pulsars that are produced.
However, one of the advantages in coupling pulsar evolution with a
full binary evolution algorithm is that it allows us to quantify the
range of initial binary parameters from which pulsars and 
particularly MSPs are generated.
It also faciliates the investigate of how uncertain assumptions in binary
evolution affect these ranges \textit{and} the nature of the pulsar 
population.

\begin{figure}
  \includegraphics[width=84mm]{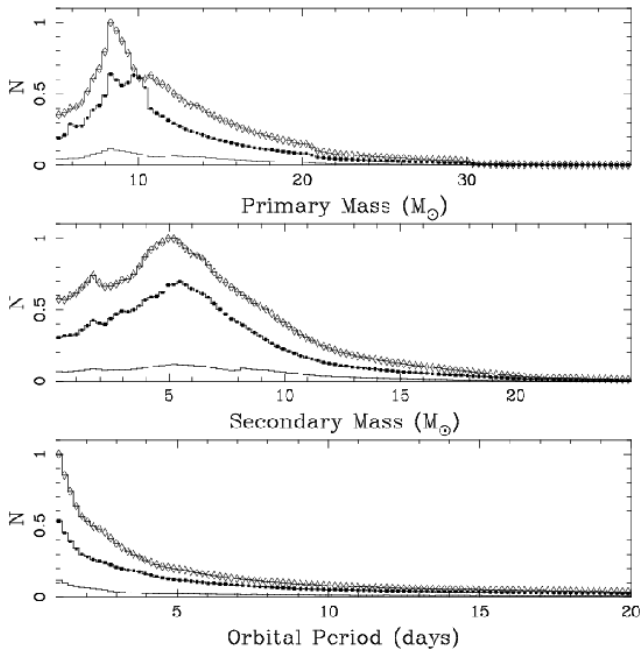}
  \caption{
Pulsar primordial parameter space distributions for primary mass,
secondary mass and orbital period.
The full squares represent the distribution of Model B, 
diamonds Model Fb and the plain line Model Fc.
Distributions are normalised to the maximum of each parameter
across the three models.
  \label{f:primrange}}
\end{figure}

In Figure~\ref{f:primrange} we show the distributions of 
initial binary parameters that lead to pulsar production in 
Models B, Fb and Fc.
In these distributions we consider only those pulsars that are plotted
in their respective model $P\dot{P}$ diagrams -- that is only those pulsars
above the death line and for Model Fc those pulsars that satisfy the 
assumed selection effects.
Also, a binary system is only considered once, even if that system produces
two pulsars in the observable $P\dot{P}$ region 
(although this is relatively rare).
We also include pulsars that lie on the lower magnetic field limit.
All three primary mass distributions peak at around $8~M_\odot$.
This value is the minimum mass in which a star is guaranteed to form a NS
(based on the stellar evolution algorithm).
As we have addressed previously (Section~\ref{s:ecsn}), 
the production of a NS from stars in the $\sim 6 - 8~M_\odot$ 
primary mass range will
depend upon the evolutionary pathway, particularly any 
associated mass loss/gain.
The drop off at around $11~M_\odot$ in the mass distributions 
is chiefly a stellar evolution feature
-- for solar metallicity models stars with $M>11~M_\odot$ ignite
core-helium burning before the giant branch which gives a change in the
radius evolution and stellar lifetime.
Formation of NSs with primary masses less than $6~M_\odot$ can only
occur via mass accretion or coalescence.
The difference in the peak in and around $8~M_\odot$ between Model B and
Fb is due to propeller evolution.
Those NSs that accrete mass and go on to collapse further into 
BHs, in Model B, instead accrete less material due to the propeller
phase in Model Fb and thus live longer lives as NSs and as such are more 
likely to be `observed' within \textsc{binpop}.
Although not examined in much detail here, at first glance one 
may consider this process to have a highly detrimental 
effect on the birth rate of BH low mass X-ray binaries when
compared to NS low mass X-ray binaries (see Kiel \& Hurley 
2006 for discussion).
However, because the NS is in a propeller phase for up to hundreds of
millions of years -- and will not be counted as a 
NS low mass X-ray binary during this time -- these systems will not add  
greatly to the birth-rate calculations of these X-ray systems.
In comparison to Models B and Fb, Model Fc producess a lower 
number of pulsars, as expected.
The distribution of secondary masses for all three models peaks at around 
$5~M_\odot$ and covers the range from $0.1 - 33~M_\odot$.
This peak is due to a combination of the lower limit of $5~M_\odot$
for primary stars, the fact that dynamic or long lived mass transfer events
occur most often in binary systems in which the mass ratio is near unity 
and that the coelescence of two $\sim 5~M_\odot$ stars leaves one 
that is able to form a NS.
Finally we come to the orbital period distribution where we find that 
short orbital periods are favoured.
Bearing in mind that the initial distribution of periods is flat in 
$\log{P}$, and thus, the relative number of binary systems born with 
periods in the $1-10$ d and $10-100$ d ranges are the same, 
it is interesting to see that the period distribution of pulsar progenitors 
does not follow that of all binaries.
This indicates that close binary evolution is instrumental in increasing
the number of observable pulsars.
With the advent of detailed pulsar evolution in combination with 
rapid binary evolution it is now possible (as demonstrated here) 
to consider the underlying main sequence stellar population 
that generates the complete observed radio pulsar distribution.

\subsubsection{Observational constraints and predictions}
\label{s:further}

\begin{figure}
  \includegraphics[width=84mm]{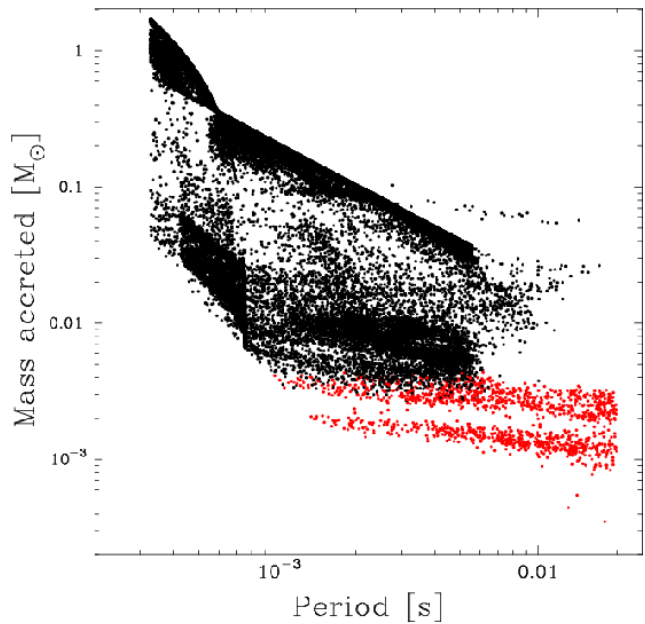}
  \caption{
Mass accreted since SNe for all pulsars rotating with 
spin periods faster than $0.02$ s excluding those that
fail our energy conservation check (see text for details).
These pulsars are formed from Model Fd (with out propeller
physics included).
Black points represent those unobservable pulsars whose 
magnetic field has decayed to the assumed bottom magnetic
field limit.
All other points represent potentially observable MSPs.
  \label{f:MSPmassacc}}
\end{figure}

Conclusions on how complete our knowledge truly is for aspects of 
binary evolution theory may be drawn if we compare certain pulsar 
population features directly with observations.
This may be completed in some limited form even though we do not yet
consider selection effects fully.
To do this we explore in greater detail the accretion physics at play 
in Model Fd.
In particular, we examine MSP mass accretion showing that under standard
accretion physics assumptions the RLOF-induced MSPs
accrete enough material for their magnetic fields to decay beyond 
observed values (in this model we assume a bottom field of $5\times10^7$ G).
Figure~\ref{f:MSPmassacc} shows that many MSPs that were spun-up via
RLOF accreted greater than $\sim 0.1~M_\odot$ 
worth of mass.
Which, according to the prescription presented here is enough 
accreted material to decay/bury the field.
In fact we find that $\sim 0.004~M_\odot$ is enough mass for a 
NS to accrete and sit on or close to the bottom field, or if we 
assume no bottom field it is enough mass that the NS will not 
be observable as a pulsar.
We note that the MSP spin periods calculated in this 
work should be considered lower limits.
The \textsc{bse} algorithm takes care to conserve angular 
momentum and the spin period resulting from the accretion of 
material is based on the angular momentum transfered by the material.
We also perform a check that this does not violate energy conservation 
(which affects the lower left-hand region of Figure~\ref{f:MSPmassacc}),
however, this check currently assumes that all the gravitational 
potential energy of the accreted mass is converted into 
rotational energy for the accreting star.
Refining this treatment will be investigated further in the future.

\begin{figure}
  \includegraphics[width=84mm]{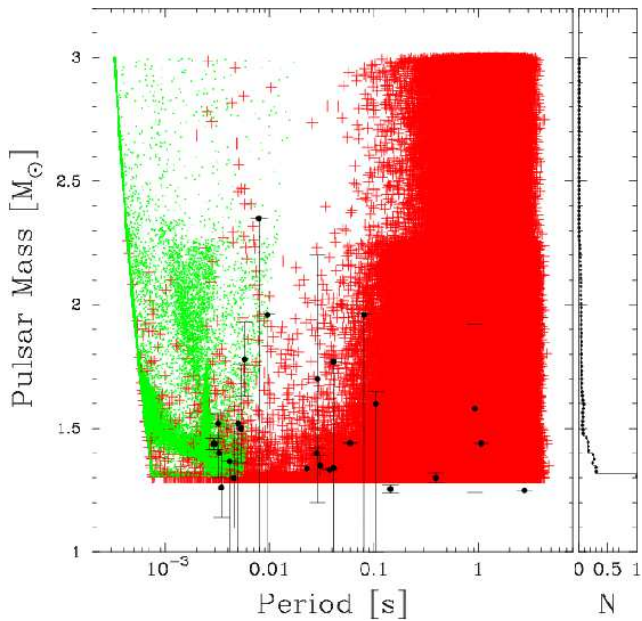}
  \caption{
NS mass versus spin period for the complete population
of Model Fd.
Plus symbols represent NS formed from our model with
$B_{\rm s} > 6\times10^7~$G (and thus potentially observable)
while the light dots represent all other model pulsars.
The darker points show observations and their suggested
errors.
On the right we show the density of pulsar mass.
We see that the majority of pulsars have masses 
in the range $1.3$ to $1.5~M_\odot$.
  \label{f:Pulsarmasses}}
\end{figure}

We also consider the mass range of pulsars produced in the above suggested 
model and compare this with observations.
Figure~\ref{f:Pulsarmasses} depicts the distribution of NS
mass against spin period and provides the number density of pulsars 
over the entire mass range.
Note that in our models we assume a maximum NS mass of $3~M_\odot$,
which explains the upper cut-off in the mass distribution (and also
affects how much mass a NS can accrete before becoming a BH).
Overlaid on our model mass distribution is the observed 
pulsar masses with their errors as quoted from Table 2 of Nice (2006;
and references therein) and Table 1 of Freire (2007; except for 
pulsar J1748-2021B).
We also include the latest mass estimate of MSP J0437-4715
by Verbiest et al. (2007) and updates of J0705+1807, J0621+1002 and 
J1906+0746 from Nice, Ingrid \& Kasian (2008), while we make use of 
Lattimer \& Prakash (2007) for pulsars J0045-7319 and J1811-1736.
At this stage, due to the lack of observational constraints, we believe
it would be inadvisable to put forward strong views on this matter.
It is also instructive to see what occurs if we ignore those 
MSP-like NSs that would most likely be unobservable due to 
accretion induced field decay (again see Figure~\ref{f:Pulsarmasses}).
We see that a large number of the highly recycled pulsars
are now removed from the mass distribution.
When only considering those potentially observable re-cycled NSs 
the apparent overlap of observed to theoretically produced pulsars, 
in the spin period $\sim 0.003$ to $0.01$ range, is now more
prominent.
Beyond this spin range the number density of observable 
model pulsars drops off somewhat.

\begin{figure}
  \includegraphics[width=84mm]{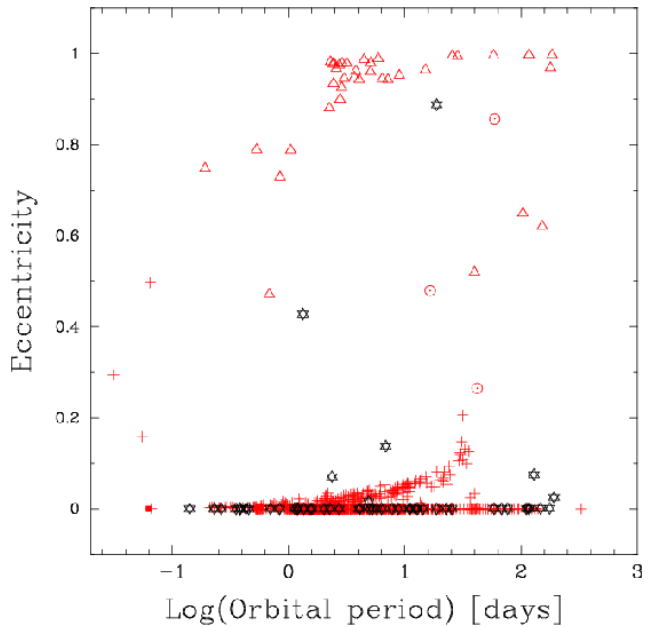}
  \caption{
MSP eccentricity versus orbital period for the MSP population
of Model Fd.
All pulsars here have magnetic fields such that 
$B_{\rm s} > 6\times10^7$ G and spin periods less $0.02$ s.
Triangles represent model double NS systems containing one MSP.
Plus symbols are WD-MSP systems, $\odot$ symbols represent
main sequence-MSP systems while solid sqares represents 
naked helium-MSP systems.
There are not many naked helium-MSPs within this figure 
and all of them reside in circular orbits, thus rendering them
hard to see.
The black stars are observed pulsars taken from the ATNF 
pulsar catalogue (Manchester et al. 2005).
  \label{f:Pulsarecc}}
\end{figure}

In keeping with our discussion on MSPs we now consider what
MSP companion types are produced by Model Fd and the ratio
of eccentric to circular orbits in MSP binaries.
MSP binary systems comprise much of the interesting binary 
evolutionary phases in their histories and as such
will play an important role in constraining binary and stellar 
evolution theory via population synthesis.
Figure~\ref{f:Pulsarecc} depicts our Model Fd MSPs in the 
eccentricity-orbital period parameter space.
The model MSPs in Figure~\ref{f:Pulsarecc} are selected to have 
$B_{\rm s} > 6\times10^7~$G and spin periods less $0.02~$s.
Model Fd is also able to produce eccentric WD-MSP systems
although the most eccentric of these systems in which the MSP is not
in the process of accreting material is one with an eccentricity
of $0.2$.
All eccentric MSP-MS binary systems depicted within 
Figure~\ref{f:Pulsarecc} are also accreting and 
thus not observable as radio pulsars.
It is interesting to note the healthy amount of double NS 
systems Model Fd produces which include a millisecond pulsar 
and a large eccentricity.
As an example to the extent of predictive power our future 
code will have on the possible understanding or interpretation of 
observations, we now consider the MSP J0514-4002A (Freire et al. 2004). 
This is the observed binary system depicted in Figure~\ref{f:Pulsarecc} 
with an eccentricity of $0.888$.
At present this binary system has a measured mass of 
$M_{\rm system} = 2.453~M_\odot$ and resides within the globular cluster
NGC 1851.
The MSP, which rotates at $P = 0.004991~$s, has an estimated mass 
of $< 1.5~M_\odot$ which leads Freire, Ransom \& Gupta (2007) to 
argue that the minimum mass of the companion is $0.91~M_\odot$.
This being the case Freire, Ransom \& Gupta (2007) suggest that the
companion is a heavy WD and that the most likely formation mechanism
is via third body interaction due to the globular cluster high 
stellar density.
Although the likelihood of this formation mechanism being 
the correct one is high they do suggest an alternative --
a near head-on collision of a pulsar and giant star.
However, because of where J0514-4002A sits within Figure~\ref{f:Pulsarecc} 
we consider a different possible binary companion type and 
formation mechanism for J0514-4002A.
Some of the highly eccentric MSPs clustered around 
$1 < P_{\rm orb} < 10~$days within Figure~\ref{f:Pulsarecc} have low 
mass primary and secondary NSs, that is $M_{\rm system} < 2.7~M_\odot$.
The MSPs within this region have accreted only $\sim 10^{-3}~M_\odot$
and have been spun-up to periods, $2\times10^{-3} < P < 10^{-2}~$s.
So, we have some low mass systems which roughly fit the observed binary 
system containing MSP J0514-4002A, bearing in mind that our minimum 
initial NS mass is $1.3~M_\odot$.
This is especially the case if one assumes a minimum NS mass 
of $\sim 1.2~M_\odot$ -- the latest minimum pulsar
mass estimate (Faulkner et al. 2005).
Prior to this it was assumed that pulsar masses were 
greater than $1.3~M_\odot$ -- the value which \textsc{bse} was 
calibrated to.
This is something that will need to be revised in the future.
Therefore, the model suggests that a low-mass highly eccentric
MSP-NS system should also be considered as a possible
explanation for this binary (in addition to the eccentric MSP-WD scenario).
We may also play this game for the Galactic disk eccentric millisecond 
binary system PSR J1903+0327 which has an orbital period 
of $95~$days and an eccentricity of $0.44$ (Champion et al. 2008).
Although not plotted in Figure~\ref{f:Pulsarecc}, this system sits close
to three double NS systems, perhaps suggesting that
PSR J1903+0327 could also be a low mass eccentric MSP-NS binary system.
The model ratio of eccentric to circular MSP systems is $0.34$.
This is assuming that any system with an eccentricity less than
$0.001$ is counted as circular.
The model does not follow eccentricity evolution
below $0.001$ owing to concerns about accuracy, i.e. systems
with eccentricities lower than this are considered to be circular.

Although not shown here directly it is possible for us to 
produce MSP-BH systems.
Unfortunately, all of the MSPs in these systems accrete
enough material that their magnetic fields decay
quickly beyond the pulsar visibility limit.
However, it is interesting to see that we are able
to produce rapidly rotating NSs with BH companions.
A typical evolutionary formation scenario begins with
two large stars, the primary $M_1 \sim 65~M_\odot$ and secondary 
$M_2 \sim 30~M_\odot$ in a close orbit separation $a \sim 40~R_\odot$ 
(an orbital period of $\sim 5$ days).
Both stars begin to rapidly lose mass in winds and eventually the primary
inititates mass transfer, at a system time of $2.5~$Myr by which time it 
has already lost $\sim 4~M_\odot$ and the companion 
has lost $\sim 0.3~M_\odot$.
When the RLOF mass-transfer phase ends some $2~$Myr
later the primary star is now a core-helium burning star 
with $8~M_\odot$ and the secondary is a $66~M_\odot$ main sequence star.
The orbit has increased to a separation of $\sim 180~R_\odot$.
The primary star then rapidly evolves through naked helium phases
to become a $2.4~M_\odot$ NS.
The system survives the SN event and is now $5.2~$Myr old with a 
separation of $\sim 400~R_\odot$ between the stars.
The companion begins to evolve quickly off the main sequence and in the 
$0.5~$Myrs it takes to become a $13~M_\odot$ BH it loses $30~M_\odot$
in a wind.
The NS has accreted $0.07~M_\odot$ of this material which 
is enough to spin it up to a millisecond spin period.
We find that $0.8\%$ of the binary pulsars rotating with periods
less than $0.02$ have a BH companion (including those that have 
decayed to the bottom magnetic field limit).
Of these rapidly rotating NSs with BH companions just under half 
occur with orbital periods less than $50~$days (the smallest
orbital period is just under $0.1~$days).
Also, the majority of them ($94\%$) have eccentricities greater than
$0.2$.
It can be supposed that if one modifies our method for 
decaying the magnetic field during accretion such that (some of) 
the above system MSPs do not decay beyond the pulsar death line, 
it could be possible to produce observable MSP-BH systems in 
relativisitic orbits (that is a heavy, tight and eccentric binary 
system in which one may constrain general relativity) within the 
Galactic disk.

\section{Discussion} 
\label{s:disc}

In the previous sections we investigated a range of 
input parameters and evolutionary assumptions that affect 
the outcomes for isolated and binary pulsar evolution.
This lead us to a preferred model which gave good coverage of the 
$P\dot{P}$ parameter space.
For that model we examined NS and orbital properties comparing
them to observations.
We next inspect some aspects of the results in further detail and
comment on areas we think are important for future work. 

Modelling beaming and accretion selection effects significantly reduces 
the number of observable pulsars predicted by our model.
Within this Model Fc there was uncertainty in how to deal with 
Thorne-\.{Z}ytkow objects.
Within Section~\ref{s:updatedist} our assumption was that the NS reset 
during the destruction of the T\.{Z}O (made for simplicity).
At present we are unsure if this is a realistic assumption to make.
If we instead assume the pulsar is unchanged when passing through a
T\.{Z}O evolutionary phase we find many more isolated 
MSP systems are produced.
Coupled to this, in Section~\ref{s:Beam}, we assumed that 
when the material is expelled from the T\.{Z}O, leaving 
behind a NS, that the expelled material enshrouds the NS 
for long enough that it is not observed before crossing 
the death line (very much like the assumption for post-accretion NSs).
Again, we are unsure how physically correct this assumption is.
Below are a number of other possible evolutionary assumptions that 
could be made when dealing with these systems.
Perhaps the NS should accrete some amount of material, decaying the 
magnetic field, and, if enough material is accreted, form a BH 
(Fryer, Benz \& Herant 1996).
Or if a BH is not formed perhaps a revitalised NS is born 
(Podsiadlowski 1996) -- a standard isolated pulsar, 
or a pulsar which is below the death line.
But one may even consider an outcome that is quite interesting.
That is the possible formation of a magnetar.
It could even be assumed that the surrounding material causes a severe
braking mechanism, the outcome of which is a young hot highly magnetic
slowly-rotating NS (see below for further discussion on possible
magnetar population synthesis).

We showed in Section~\ref{s:Mod2} that values of the magnetic field
decay timescale, $\tau_{\rm B}$, below $100~$Myrs do not provide
a realistic distribution of pulsars.
Then in Section~\ref{s:updatedist} we explained that using
$\tau_{\rm B} = 1000~$Myr does not cover enough of the slow 
rotating pulsars in the $P\dot{P}$ diagram but that using
$\tau_{\rm B} > 2000~$Myr over corrected for this.
For our preferred model we found $\tau_{\rm B} = 2000~$Myr to be optimal.
We note that the choice of maximum age for the population, in our
case $10~$Gyr, affects the time available for decay and as a result
affects the appearance of the pulsar distribution at the lower edge
of the main pulsar island.
However, the appearance of this region also depends crucially on the
details and/or shortcomings of the assumed death line used in the
model (which we now discuss).

In all our models the recycled pulsars are not suitably limited by
the death line.
Of course, the inverse Compton scattered death line models have not been
considered in any detail, and as Hibschman \& Arons (2001) suggest all three
models may be required to suitably simulate the complete pulsar 
population death line.
In fact, Hibschman \& Arons (2001) show regions of $P\dot{P}$ in which
the different methods would come into play.
The standard pulsar island is shown to be modelled by CR, the recycled
pulsars by nonresonant inverse Compton scatter and the high magnetic 
field pulsars governed by the resonant inverse Compton scattered method.
This matches our findings that CR models the standard pulsar island death 
line in a suitable manner but is too lax in culling recycled pulsars.
To test this further we now examine what effect decreasing 
$\bar{f}^{\rm min}_{\rm prim}$ to $0.10$ (from the previously assumed $0.15$) 
has on our pulsar population.
The recycled pulsar island is now more effectively 
constrained by the death line.
However, the death line is now too effective at culling pulsars from 
the standard pulsar island.
As explained previously (see section~\ref{s:death}), the death line is 
broken up into two regimes, saturated and unsaturated.
Within all the models so far it seems that the majority of pulsars
are constrained by the saturated death line region (even if we are to
decrease $\bar{f}^{\rm min}_{\rm prim}$ to $0.03$).
In fact, any pulsar at the begining of its accretion phase is governed 
by the saturated regime (although at some point after this it may move 
into the unsaturated regime for a time).
To increase the influence of the unsaturated regime we can
take some liberties with the assumed unsaturated 
$\bar{f}^{\rm min}_{\rm prim}$ value and let it approach zero. 
However, this eliminates a wedge of pulsars from a region of the 
$P\dot{P}$ space in which pulsars are observed and this is not an 
acceptable parameter change.
We note that death line calculations are model dependent and in 
particular the assumed NS equation of state is an important feature 
and is something we have not considered here in any detail.
Further work is required in examining all the Harding, Muslimov \& 
Zhang (2002) death lines but is beyond the scope
of this paper.

All the models presented here allow the creation of NSs with
spin frequencies in excess of $1000~\rm{Hz}$.
The only reason these NSs do not spin beyond $\sim 3300\rm{Hz}$
(see, say, Figure~\ref{f:res8}) is that, within \textsc{bse},
no star is allowed to spin beyond their break-up speed.
There are theoretical mechanisms that allow the regulation 
of pulsar spin and restrict it to some spin velocity less than the break-up 
velocity.
One such possible mechanism is gravitational emission from the
pulsar itself.
There are two theories as to how a pulsar may emit gravitational 
radiation and thus self-regulate its own spin period.
One theory suggests that if a pulsar surface is not smooth or purely 
spherical (this would be the case for those that have accreted 
material or that have strong magnetic fields and are rotating rapidly)
then this asymmetry in the pulsars features may be mapped to the 
pulsars space time curvature and cause energy disipation in the 
form of gravity waves (Shapiro \& Teukolsky 1983; Bildsten 1998;
Payne \& Melatos 2005).
The other theory assumes that the Rosby wave (r-wave or r-mode) 
instability (Levin 1999; Friedman \& Lockitch 2001) is not damped.
This instability basically arises due to the form of the wave and 
the speed of rotation of the pulsar.
Chandrasekhar (1970) showed that Maclaurin spheriods\footnote{A method
for modeling uniform density ellipsoids. In particular it gives a 
description of the relationship between the eccentricity of the system,
in this case the pulsar, and its uniform angular momentum.}
with uniform density, rotating uniformly are unstable to 
$\rm{l} = \rm{m} = 2$ polar modes (i.e., those waves in which the 
wave front stretches from pole to pole and travels around the sphere).
In a non-rotating or slow rotating star, a particular mode, 
the non-axisymmetric instability removes positive and negative 
angular momentum from the forward and backward mode respectively 
(Friedman \& Lockitch, 2001), thus all the non-axisymetric modes 
cancel out and no instability forms.
In a rapidly rotating star ($\sim$ that around the maximum observed
pulsar spin rate; Levin 1999), a backward moving mode may be moving 
forward as observed by an inertial observer and therefore radiate away 
positive angular momentum -- in effect increasing the mode amplitude 
(Friedman \& Lockitch 2001).
In this manner gravitational radiation drives the mode and the star 
spins down until the instability dies (Levin 1999).

With the suggested discovery of a pulsar with sub-millisecond period (Kaaret 
et al. 2007), the maximum spin rate possible for a pulsar is work we 
would like to consider in the future.
At the forefront of this work will be contemplation of the above
gravitational radiation pulsar spin-down methods.
However, we wish to point out that the assumed accretion-induced
magnetic field decay parameter may play an important role in determining
the observed MSP spin period cut off (see Section~\ref{s:accres})
and is something that will need to be considered in future work.

In recent years magnetars have become a distinct subclass of the 
pulsar population.
These systems are highly magnetic NSs with magnetic fields of order $10^{14}
- 10^{15}$ G.
Magnetar formation is not well understood, though work 
by Thompson \& Duncan (1993) and Duncan \& Thompson (1992)
showed that nascent NSs which rotate rapidly ($\la 0.03$ s) can have magnetic
fields within the above range.
In terms of the work being completed here we have not yet considered 
the possibility of modeling these highly magnetised, slow rotating pulsars.
The reasoning behind this is that we are trying to produce radio pulsars
and these systems have only been found in X-rays (anomalous X-ray 
repeaters) and gamma-rays (soft gamma-ray repeaters:
see Woods \& Thompson 2006 for review).
However, in light of the now constant flow of new and improving
theoretical and observational considerations of these objects
(Spruit \& Phinney 1998; Heger et al. 2003; 
Heger, Woosley \& Spruit 2005; Mezzacappa 2005; Thompson 2006; 
Harding \& Lai 2006; Blondin \& Mezzacappa 2006) an attempt 
to model these systems with very simple assumptions is now possible 
and will be included in future work.

\section{Conclusions}
We model the evolution of radio pulsars, both single 
and in binary systems. 
Models are evolved from realistic initial mass functions, 
period distribution and parameter ranges.
Our primary goal is to demonstrate that we have an evolution algorithm
that can generate the range of pulsars observed in the Galaxy,
thus laying the groundwork for a synthetic Galactic population code
that will ultimately include kinematics and selection effects.
However, through investigation of the $P\dot{P}$ diagram (comparing
model-model and model-observation) we have already been able to 
demonstrate how uncertain factors of pulsar and binary evolution 
can be constrained.
We have also been able to make preliminary predictions about the Galactic
pulsar population, in particular MSPs.
This represent the first study combining detailed binary evolution 
and pulsar physics. 
Of course a study such as this is hampered by many poorly 
constrained parameters which may result in misleading 
results.
This is especially true if the complete relevant parameter 
space is not examined and multiple observations are not used 
as benchmarks.
Therefore, we once again remind the reader that some caution is 
required when examining our results.

In terms of pulsar evolution it has been possible to provide 
constraints on various aspects of the modelling.
For instance, our models can demonstrate that magnetic field decay 
timescales of order $100~$Myr or less can not produce the required 
distribution of spun-up pulsars.
We also show that propeller evolution has a large impact on pulsar 
populations and can be an important component of pulsar modelling.
In particular we show that the inclusion of the propeller mechanism
causes wind accretion to be the dominant pathway for MSP production.
In fact, neglecting propeller evolution is necessary in order to 
produce any sort of RLOF disk accretion powered MSP
(the typical type of observed accretion powered MSP).
To further investigate this aspect there is a need to examine methods for
dealing with accretion physics in more detail -- including the addition of
transient RLOF accretion features.
We find that accretion-induced magnetic field decay (or apparent field decay) 
provides a natural method for forming rapidly rotating low magnetic 
field NSs.
Without any such magnetic field decay we find there is no hope of 
theoretically reproducing the observed pulsar $P\dot{P}$ distribution.
The accretion-induced method assumed here works well for all models 
performed and we depict how changes in the efficency of the decay
modifies the populated $P\dot{P}$ region.
Including an angular momentum wind accretion efficency parameter
-- lowering the angular momentum accreted -- 
assists in producing some of the tightest observational $P\dot{P}$ 
constraints.
We also find that the addition of an electron capture assymetric SN kick 
distribution results in the formation of a greater number of MSPs
than in previous models.
As such this evolutionary assumption will doubtlessly be an important 
feature in future works and requires further examination.

We demonstrate the predictive power of our code by examining 
the pulsar mass range and eccentricity of MSP 
systems produced in a favoured model.
We must wait to impose selection effects before we can advise as to
what fraction of these are potentially observable.
We also show that under standard accretion physics assumptions NSs are 
able to accrete up to $\sim 1~M_\odot$ of material, which can spin them 
up to rapid rotation rates and decay their magnetic fields to strengths 
considered insufficient for producing radio emission.

We conduct an initial investigation into how the $P\dot{P}$ parameter 
space is modified when selection effects are considered.
These include beaming and blanketing of the pulse by material thrown off 
by stellar winds or during mass accretion events.
We study the distributions of primordial primary mass, secondary mass and 
orbital period for binaries that produce pulsars.
Changes in these distributions due to assumed evolutionary variations are 
examined.
In particular we show that close binary evolution is effective in 
boosting the numbers of observable pulsars.
Finally, it is clear from all of our models that a greater level
of detail is required when calculating the complete pulsar
population death line.
 
The work shown here makes use of the first of three modules that 
will eventually be able to simulate pulsar observational surveys, 
including next generation radio surveys such as the Square 
Kilometer Array. 
This involves following the positions of pulsars within the 
Galactic gravitational potential and modelling survey selection 
effects. 
The long term aim is to include multi-wavelength survey capability 
into our models, and thus observe any Galactic stellar population.
It is important to point out here that once complete this 
code will consist of the most comprehensive treatment of 
input physics and selection effects for pulsar 
simulation, something not previously attempted.

\section*{Acknowledgments}
PDK wishes to thank Swinburne University of Technology for a
PhD scholarship.
PDK also thanks Joris Verbiest for discussions on observed pulsar masses.
The authors wish to thank the referee who helped improve the
focus of this work.

\bsp

\label{lastpage}

\end{document}